\documentstyle[12pt,psfig]{article}

\begin{document}
\begin{flushright}
TRI-PP-95-53\\
PSI-PR-95-14\\
\end{flushright}
\vspace{1cm}
\begin{center}
{\large \bf First Order Variational Calculation of Form Factor
in a Scalar Nucleon--Meson Theory}

\vspace{1cm}
A.~W.~Schreiber

\vspace{0.2cm}
TRIUMF, 4004 Wesbrook Mall, Vancouver B. C., Canada V6T 2A3

\vspace{0.5cm}
R.~Rosenfelder

\vspace{0.2cm}
Paul Scherrer Institute, CH-5232 Villigen PSI, Switzerland

\end{center}

\vspace{1cm}
\begin{abstract}
\noindent
We investigate a relativistic quantum field theory in the particle
representation using a
non-perturbative variational technique.  The theory is that
of two massive scalar particles, `nucleons' and `mesons',
interacting via a Yukawa coupling.
We calculate the general Euclidean Green function involving two
external nucleons and an arbitrary number of external mesons in
the quenched approximation for the nucleons.  The non-perturbative
renormalization and truncation is done in a consistent manner and
results in the same variational functional independent of
the number of external mesons. We check that the calculation
agrees with one-loop perturbation theory for small couplings. As an
illustration the special case of
 meson absorption on the nucleon is considered in detail. We derive
the radius of the dressed particle and
numerically investigate the vertex function after analytic continuation
to Minkowski space.
\end{abstract}

\vspace{1cm}
\newpage

\section{Introduction}

\noindent
There are many instances in relativistic quantum field theory where
perturbation theory is inadequate, in particular when the coupling
constant characterizing the interaction becomes too large.
An alternative approach, which remains valid in
the strong coupling regime, is the use of a variational principle.
Although there exists a sizable body of literature on the use of
variational methods in relativistic quantum field theories~
\cite{Sch,CJT,Stev,Fey2,Hay} it is
fair to say that these methods are less widely used in this area of
physics than the standard Rayleigh-Ritz variational method in
non-relativistic quantum mechanics.
This is largely due to the fact that one has to deal with a system
with infinitely many degrees of freedom, the requirements of
relativistic invariance and  renormalizability
of the theory.

The problem of the infinitely many degrees of freedom, however, is
also present in non-relativistic field theories and
here there exists a variational method which deals with this problem
very successfully -- Feynman's variational treatment of the
polaron~\cite{Fey1,FeHi,RoFe,MCM,BoPl,GeLo,AlRo}.  A very early
attempt to use this method in a relativistic setting was made in
Ref.~\cite{Mano} and was recently reinvented and extended by
us~\cite{RS1,RS2,SRA}. (The method has also previously been adapted
for use in non-relativistic nuclear
physics~\cite{Pol1,Pol2,AlDi}).  The particular Lagrangian
which was considered in these papers was that of two scalar
particles, one
heavy (henceforth referred to as the `nucleon') and one light
(the `meson'), interacting via a Yukawa
coupling. This variant of the Wick-Cutkosky model~\cite{Wick,Cut}
was chosen because of its simplicity, its similarity to several
realistic Lagrangians (such
as those of QED and nucleon-meson theories) and
because it has traditionally been used for the study of
non-perturbative techniques~\cite{SiTj,NTS,LeDa,Saw,Hil,WiHi,Ji}.

The quantities of interest in a relativistic field theory are the
Green functions of the
theory.  The work in Refs.~\cite{RS1,RS2,SRA} concerned itself with
the calculation of a subset of these Green functions in the
quenched approximation for the nucleon. In particular,
Green functions with 2 external (on-shell) nucleon legs and any
number of external meson legs were calculated in Ref.~\cite{SRA}.

Feynman's variational method involves the use of the path integral
formalism.  As is well known, in general it is
only possible to do gaussian path integrals analytically.  Because
of this he made use of a particular approximation --
Jensen's inequality which can be exploited for variational purposes.
Equivalently, this
corresponds to the first order term in a cumulant expansion
(see, for example, Ref.~\cite{KLB}). In Refs.~\cite{RS1,RS2}, the
variational  calculation of the nucleon propagator was performed
using the same approximation.
The only quantity of interest there was the residue of this
propagator, the position of the pole being defined as the
(non-perturbatively) renormalized mass of the nucleon.
(In addition, the `width' of the nucleon was calculated --
this quantity is not zero as the Wick-Cutkosky model is
inherently unstable.) In Ref.~\cite{SRA} the general
$(2 + n)$-point function was then written down to
zeroth order in the cumulant expansion.  Although already
{\it qualitatively} interesting at this order
(for example, even at zeroth order these Green functions contain
parts of all Feynman diagrams to all orders in the coupling),
these Green functions clearly had some unwanted {\it numerical}
features.  In particular, at this order all dependence on
the external momenta is gaussian, resulting in, for example, a
gaussian form factor.
Furthermore, one of the strong points of Feynman's treatment of
the polaron was that agreement with perturbation theory was assured
to the order in the cumulant expansion in
which one calculates.  Hence, the $(2 + n)$-point functions
calculated in Ref.~\cite{SRA} only
agree with perturbation theory at tree level.

In the present paper we calculate the $(2 + n)$-point function to
first order in the cumulant expansion, therefore greatly
improving the numerical results.  More importantly,
it will be shown that the variational equations for any of these
Green functions are in fact the same, i.e. once one has determined
the numerical value of the variational parameters from one of these
functions, say the propagator of the nucleon ($n$ = 0),
the other ones may be calculated with the use of the
same parameters. It will also be shown that it is possible to
truncate the $(2 + n)$-point function consistently; that is,
the propagators on the external nucleon legs have poles at the
physical nucleon mass, as they should.

An important ingredient of the method has been the use of the
particle representation of a field theory
\cite{FeHi,Fock,Nambu,FeyQED,Schwi}, which has undergone
something of a revival in recent years
\cite{SiTj,NTS,KaKt,McRe,Str,ScSc,TS}.
We start in  Section~\ref{sec: particle representation} by
briefly outlining how one derives the expression for the quenched
$(2 + n)$-point function in the particle representation.
Having done this, we perform the actual variational calculation
in Section~\ref{sec: formalism}, resulting in the `master formula'
for the $(2 + n)$-point function
at first order in the cumulant expansion.  An important check, that
this formula agrees with perturbation
theory to one-loop order, is relegated to the Appendix.  It should
be noted that throughout this derivation we work in Euclidean
space-time.  The Green functions can however be continued to
Minkowski space.  This is done, for the $(2 + 1)$-point function,
in Section~\ref{sec: Vertex} and numerical results are derived and
discussed in Section~\ref{sec:Discussion}.
We conclude with some general observations in
Section~\ref{sec:Conclusion}.

\section{The $(2 + \lowercase{n})$-point function in the particle
representation}
\label{sec: particle representation}

\noindent
Before we turn to the calculation of the general quenched Green
function with two
external nucleon and $n$ external meson legs in first order
in the cumulant expansion, we need to derive the particle
representation of this
function.  For the propagator this was already done in
Ref.~\cite{RS1}, so here we shall limit ourselves to
briefly summarizing the relevant steps involved.
We shall follow the notation of Ref.~\cite{RS1}, and the reader
who is
interested in the details is referred to this reference.
The Wick--Cutkosky Lagrangian is given, in Euclidean space-time, by
\begin{equation}
{\cal L} = \frac{1}{2} \left ( \partial_{\mu} \Phi \right )^2 +
\frac{1}{2}
M_0^2 \Phi^2 + \frac{1}{2} \left ( \partial_{\mu} \varphi \right )^2
 + \frac{1}{2} m^2 \varphi^2 - g \Phi^2 \varphi
\end{equation}
where $M_0$ is the bare mass of the heavy nucleon, $\varphi$ and
 $\Phi$ are the fields of the meson and nucleon respectively
and $m$ is the meson mass.  The latter does not get renormalized
in the quenched approximation while the nucleon mass does.
  The corresponding generating functional for the general
Green function is written in terms
of the sources $J$ and $j$ of the external nucleons and mesons as
\begin{equation}
Z\> [J,j] = \int {\cal D} \Phi  \> {\cal D} \varphi \>
\exp \left ( - S[\Phi,\varphi]
+ (J,\Phi) + (j,\varphi) \> \right )\;\;\;.
\label{generating}
\end{equation}
Here the action is given by
\begin{equation}
S[\Phi,\varphi] = \int d^4x \> {\cal L}(\Phi(x),\varphi(x))
\end{equation}
and we have defined
\begin{equation}
(J,\Phi) \equiv \int d^4x \> J(x) \Phi(x) \>\>\>\> {\rm etc.}
\end{equation}
One implements the quenched approximation by integrating out the
nucleon field and setting the resulting determinant equal to unity.
  After functionally differentiating twice with respect to the
nucleon source $J(x)$ and then setting this source to zero we
obtain the generating functional for the
Green functions with two external nucleon legs and an arbitrary
number of meson legs.  It is given by
\begin{eqnarray}
Z'\> [j,x] &\equiv& \frac{\delta^2 Z \> [J,j] }{\delta J(x) \>
\delta J(0)} \Biggr |_{J=0}
\label{generating' field}\\
&=& \int {\cal D} \varphi \> < x \> | \frac{1}
{-\Box + M_0^2 - 2g \varphi} |\> 0 >
\> \exp \left [ -\frac{1}{2} (\varphi, (-\Box + m^2) \varphi ) +
(j,\varphi) \right ] \> .\nonumber
\end{eqnarray}
This generating functional is written in terms of the meson field
$\varphi$.  In order to implement the particle representation of
the theory in terms of the path of the nucleon we exponentiate the
propagator  and write the resulting relativistic matrix element in
terms of a proper time path integral~\cite{Schul}:
\begin{equation}
< x \> | \exp\left [ - \beta \> (\frac{\hat p^2}{2} - g \varphi(x) )
\right ] | \> 0 >   =
\int_{x(0)=0}^{x(\beta)=x} {\cal D}x(\tau) \> \exp \left (
- \int_0^{\beta} d \tau \> \left [ \> \frac{1}{2} \dot x^2 -
g \varphi(x(\tau) ) \> \right ]
\right ) \> .
\label{coordinate path integral}
\end{equation}
We have given the particle the ``mass'' 1
which corresponds to the `proper time gauge' \cite{BDH}. Our
variational results can be shown to be gauge-independent although
some variational parameters are not \cite{RS2}.
One may now perform the functional integration over the meson
fields to obtain
 the generating functional of connected Green functions
\begin{equation}
Z'_{\rm conn}\> [j,x] = {\rm const} \int_0^{\infty} d \beta \>
\exp \left (- \frac{\beta}{2} M_0^2 \right )
\> \int_{x(0)=0}^{x(\beta)=x} {\cal D}x(\tau) \>
\exp ( - S_{\rm eff}\> [ x(\tau),j]\>  \> ) \;\;\;.
\label{generating' connected}
\end{equation}
Here the effective action is given by
\begin{equation}
S_{\rm eff}\> [x(\tau),j] = S\> [x(\tau)] \>
- g \int d^4 y \> j(y) \> \int_0^{\beta} d \tau \>
< y | \> {1 \over -\Box + m^2} \> | x(\tau) >
\label{eff action}
\end{equation}
with
\begin{eqnarray}
S\> [x(\tau)] &=& S_0\> [x(\tau)] + S_1\> [x(\tau)]
\label{action definition}\\
S_0\> [x(\tau)] &=& \int_0^{\beta} d \tau  \> \frac{1}{2} \dot x^2
\label{eff action kin}\\
S_1\> [x(\tau)] &=&
- \frac{g^2}{2} \int_0^{\beta} dt_1 \> \int_0^{\beta} dt_2 \>
< x(t_1) \left |  \> {1 \over -\Box + m^2} \> \right | x(t_2) >
\label{eff action pot}
\label{eff action disconnected}
\end{eqnarray}
Finally, one may obtain the Green function with $n$ external meson
legs by differentiating $n$ times with respect to the source $j$
and then setting this source to zero.  After Fourier transforming to
momentum space and removing the
external meson legs as well as the overall momentum conserving
delta function one obtains
\begin{eqnarray}
G_{2,n}(p,p';\{q\})&=& {\rm const.}  \int_0^\infty d\beta
\exp \left [ - {\beta \over 2} M_0^2 \right ]
\int d^{4}x \> e^{- i p'\cdot x} \nonumber \\
&&\hspace{1cm} \cdot \>
\int_{x(0)=0}^{x(\beta)=x} {\cal D} x(\tau)
\prod_{i=1}^{n} \left [ g \int_{0}^{\beta} d\tau_i
e^{i q_i \cdot x(\tau_i)}\right ] e^{-S[x(\tau)]}\;\;,
\label{defeq}
\end{eqnarray}
Here the definition of the meson momenta \{q\}  and of the final
nucleon momentum $p'$ is such that the latter is in--going while the
former are out--going. The integration parameters $\tau_i$ have an
obvious physical significance -- they mark the (proper) times at
which the external mesons interact with the (bare) nucleon.

Note that the external nucleon legs are still contained in this
expression. As the nucleon mass needs to be renormalized their
truncation is not as trivial as that of the meson legs.
In particular, the nucleon mass is a {\it dynamical} quantity
which needs to be {\it calculated} in terms of the bare nucleon mass.
  Of course, if one could calculate the Green functions of the
theory exactly, the poles of the external nucleon legs would all
be automatically at the same position as the
pole of the propagator.  In an approximate treatment, however,
this is not automatically fulfilled.  In fact, it is  non-trivial
to even isolate these poles, let alone to have them occur at the
correct position.  As will be
seen in the next Section, it is indeed possible to perform the
truncation of the general Green function in a consistent manner
within the context of the variational calculation described here.
The result (see (Eq.~\ref{eq:truncated}) ) will be
a rather compact expression for the general truncated $(2 + n)$ - point
function, which encapsulates, as we shall see, information
contained in any order of perturbation theory
as well as genuinely non-perturbative effects.

\section{Variational Calculation of the $(2 + \lowercase{n})$-point function
to first order in the cumulant expansion}
\label{sec: formalism}

\noindent
The variational treatment is based on the decomposition of the
action into a trial action $\tilde S_t$ plus a remainder
\begin{equation}
\tilde S = \tilde S_t \> + \> \tilde S \> - \tilde S_t \> =
\tilde S_t \> + \Delta \tilde S
\label{decomp}
\end{equation}
and uses the cumulant expansion to first order (which is related to
Jensen's inequality):
\begin{equation}
< e^{- \Delta \tilde S} >_{\tilde S_t}  \> \>
\simeq \> \> e^{-< \Delta \tilde S >_{\tilde S_t} } \>.
\label{Jensen stat}
\end{equation}
Here the action $\tilde S[x]$ is related to $S[x]$ through
\begin{equation}
\tilde S[x] = S[x(\tau)] + i p'\cdot x -
i \sum_{i=1}^{n} q_i \cdot x(\tau_i)
\label{tilde s}
\end{equation}
and thus contains all the dependence on the nucleon's path in
Eq.~(\ref{defeq}). The first order cumulant has the particular
property that it is stationary (for real actions even minimal)
under arbitrary variations of $ \tilde S_t$.
We shall perform the variational
calculation in momentum space so that the averages
\footnote{These averages in momentum space were denoted by $<< .. >>$ in
Ref.~\cite{RS1} in order to distinguish them from averages in
coordinate space which were indicated by a single
bracket $< .. >$.  In the present paper we shall only deal with
momentum space averaging, so as no confusion can arise we shall
simply make use of the single brackets.}
in Eq.~(\ref{Jensen stat}) are defined by~\cite{RS1}
\begin{equation}
< {\cal O} >_{\tilde S_t} \> \equiv \frac{
\int {\cal D}{\tilde x} \> {\cal O} \> \> \exp (- \tilde S_t ) }
{\int {\cal D}{\tilde x} \> \exp (- \tilde S_t )}
\label{p averaging} \>.
\end{equation}
where
\begin{equation}
\int {\cal D}{\tilde x} = \int d^4 x \int_{x(0)=0}^{x(\beta)=x} {\cal D}x
\end{equation}

\begin{figure}
\begin{center}
\begin{picture}(100,90)
\put(25,-22){\makebox(100,90){\psfig{figure=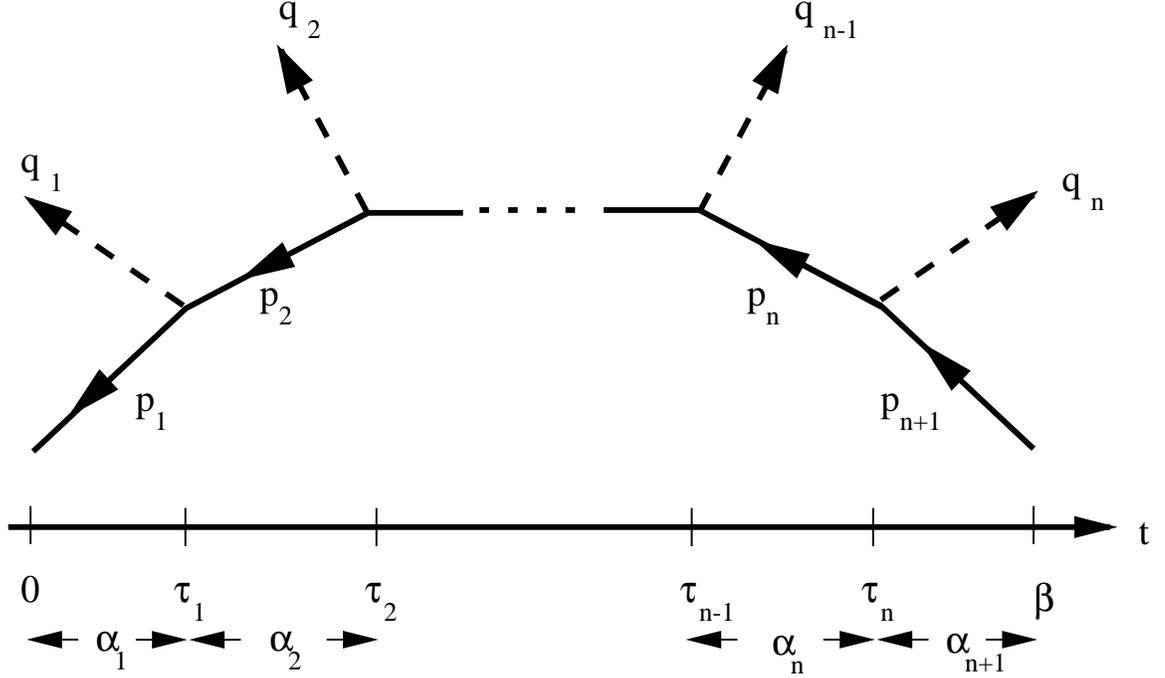}}}
\end{picture}
\end{center}
\caption{The definition of the relative times and momenta. Note
that $\tau_0 = 0$, $\tau_{n+1} = \beta$, $p_1 = - p$ and
$p_{n+1}=p'$.}
\label{relative}
\end{figure}

To this order in the cumulant expansion the $(2 + n)$-point
function is  given by
\begin{equation}
G_{2,n}(p,p';\{q\})= {\rm const.} \int_0^\infty d\beta
\left [\prod_{i=1}^{n}  g \int_{0}^{\beta} d\tau_i\right ]
\exp \left [ - {\beta \over 2} M_0^2 \right ]
e^{-< \Delta \tilde S >_{\tilde S_t} }
\int {\cal D}{\tilde x} e^{-{\tilde S_t}[x(\tau)]}
\label{greenvar}
\end{equation}
The actions $\tilde S$ and $\tilde S_t$ are  expressed in terms of
the absolute times $\tau_i$.  It will at times turn out to be more
useful to define relative times $\alpha_i$, as indicated in
Fig.~\ref{relative}. It should be noted that
Fig.~\ref{relative} is  not to be understood as a Feynman diagram -
it only serves to label the interaction times of the external
mesons with the nucleon. Furthermore,
in this Figure we have ordered these interaction times such that
$\tau_1 \leq \tau_2 .... \leq \tau_n$.  It is
always possible to do this, provided we sum over all permutations
of the $q_i$'s afterwards.

\unitlength1mm

We shall do the path integrals which are involved by expanding the
general path $x(\tau)$ into its Fourier components:
\begin{equation}
x(\tau) = x \> \frac{\tau}{\beta} + \sum_{k=1}^{\infty}
\frac{2 \sqrt{\beta}}
{k \pi} \> b_k \> \sin\left ( \frac{k \pi \tau}{\beta} \right ) \> .
\label{Fourier paramet of paths}
\end{equation}
The action $\tilde S$ now becomes
\begin{equation}
\tilde S = S[x(\tau)] + i\left [ \> \sqrt {{2\over \beta}} b_0
\cdot \sum_{i=1}^{n+1} \alpha_i p_i - \sum_{k=1}^{\infty}
Q_k^{(n)}\cdot b_k \> \right ]
\label{tilde s fourier}
\end{equation}
where
\begin{equation}
Q_k^{(n)} = {2 \sqrt {\beta} \over \pi k} \sum_{i=1}^{n}\sin
\left ( {k \pi \tau_i \over \beta} \right ) q_i\;\;\;,
\end{equation}
\begin{equation}
x = \> \sqrt{ 2 \beta} \> b_0
\end{equation}
and the definition of the momenta $p_i$ is indicated in
Fig.~\ref{relative}.

The question now arises as to what the form of the trial action
$\tilde S_t$ should be. If one were to calculate all the
corrections to Eq.~(\ref{Jensen stat})
the functional form of this trial action would in principle be
arbitrary, as would be the actual numerical values of
the variational parameters.  In practice, of course, one does
not calculate the corrections, except the lowest order one
which is needed to make the variational principle work.
It is this fact that makes
the form of the variational action assume importance.  Indeed, for
best results, the trial action  should
encapsulate as much of the information contained in the true action
as possible.  In particular, in order to reproduce the lowest
non-trivial order of perturbation theory, the trial action should
reduce to the free action in the small coupling limit~\cite{RS1}.
On the other hand, it should be of such a form that the mathematical
 manipulations required become
possible without further approximations, as otherwise one looses
control over the variational properties. In the present case this
simply means that the trial action must be quadratic in the Fourier
coefficients $b_k$ so
that the path integrals may be performed analytically. One is
therefore restricted to the following general form:
\begin{equation}
\tilde S_t = S_t + i \left [ \> \sqrt {{2\over \beta}} \rho_0 \, b_0
\cdot \sum_{i=1}^{n+1} \alpha_i p_i - \sum_{k=1}^{\infty} \rho_k \,
Q_k^{(n)}\cdot b_k \right ]\;\;\;,
\label{general trial action}
\end{equation}
where $S_t$ is given by (as in Ref.~\cite{RS1}):
\begin{equation}
S_t = \sum_{k=0}^{\infty} A_k b_k^2 \> .
\label{old trial action}
\end{equation}
Note that in the limit in which the coupling goes to zero the true
action $\tilde S$ becomes
\begin{equation}
\tilde S_0 = \sum_{k=0}^{\infty} b_k^2 + i \left [ \>
\sqrt {{2\over \beta}} b_0
\cdot \sum_{i=1}^{n+1} \alpha_i p_i - \sum_{k=1}^{\infty}
Q_k^{(n)}\cdot b_k \> \right ]\;\;\;,
\end{equation}
so that $A_k$ and $\rho_k$ must go to 1 in this limit.

The above trial action ensures that upon expansion in the coupling
constant, and without having to solve the variational equations,
the one-loop perturbative result is obtained.  However, as
indicated in Section~\ref{sec: particle representation},
 we also demand that for arbitrary
coupling one  should be able to truncate the general Green function
consistently.  It is not obvious that this is possible in general.
Indeed, we have not been able to perform this truncation with the
simplest choice of $\rho_k$ being a constant independent of $k$.
On the other hand, if one chooses $\rho_k$ to be proportional to
the `profile function' $A_k$
\begin{equation}
\rho_k = \lambda \> A_k \> \> ,
\label{eq:choice}
\end{equation}
a consistent truncation becomes possible. Here
$\lambda$ is a variational parameter which will turn out to be
precisely the same quantity which already appeared in the
study of the propagator in Refs. \cite{RS1,RS2}. Actually for
$n = 0 $ Eq. (\ref{general trial action}) with the choice
(\ref{eq:choice}) is just the ansatz which was
used there for `momentum averaging' of the 2-point function.
However, for this special case it is more a convenience than
a necessity to write the variational parameter $\rho_0$ as
$\lambda A_0$ whereas for $n > 0$ the $k$-dependence chosen in
Eq. (\ref{eq:choice}) is essential.
At the present stage we have not investigated other
parameterizations of the trial action in a systematic manner.
As the parameterization of the trial action
is connected to the analytic structure of the resulting Green
functions it would clearly be a rather interesting exercise to
determine what the most general allowable $\rho_k$ is.

One may now calculate the various quantities required for the
evaluation of the Green functions in Eq.~(\ref{greenvar}). Writing
\begin{equation}
\tilde S = \tilde S_0 + S_1
\end{equation}
one obtains, up to irrelevant overall constants,
\begin{equation}
\int {\cal D}{\tilde x} e^{-\tilde S_t} = \left [
\prod_{k=0}^{\infty} {1 \over
A_k^2}\right ] \> \exp \left \{ -{\lambda^2 \over 2 \beta} \left [
A_0 P^2 + {\beta \over 2} \sum_{k=1}^{\infty} A_k
{Q_k^{(n)}}^2 \> \right ] \> \right \}
\end{equation}
and
\begin{equation}
<\tilde S_0 - \tilde S_t >_{\tilde S_t} =
{P^2 \over \beta} \left [ \lambda - {\lambda^2 \over 2} (1+A_0)
\right ] + 2 \sum_{k=0}^{\infty} \left ( {1 \over A_k} - 1 \right )
+ \sum_{k=1}^{\infty} {{Q_k^{(n)}}^2 \over 2}
\left [ \lambda - {\lambda^2 \over 2} (1+A_k)\right ] \> .
\end{equation}
For brevity we have defined the quantity $P$ to be
\begin{equation}
P = \sum_{i=1}^{n+1} \alpha_i p_i
\end{equation}
Combining these terms yields
\begin{equation}
\exp \left [ - <\tilde S_0 - \tilde S_t>_{\tilde S_t} \right ] \>
\int \tilde {{\cal D} x} \> e^{- \tilde S_t}  =
\exp \left [ \> 2 \sum_{k=0}^{\infty} \left (1-{1 \over A_k} -
\log A_k \right ) +{\lambda (\lambda -2 ) \over 2} \sum_{i=1}^{n+1}
\alpha_i p_i^2 \> \right ]
\end{equation}
The remaining  quantity which needs to be calculated is
$<S_1>_{\tilde S_t}$. It is given by
\begin{equation}
< \> S_1 \> >_{\tilde S_t} = \> - \frac{g^2}{2} \> \int_0^{\beta}
dt_1 dt_2 \> \int \frac{d^4q'}{ (2 \pi)^4} \> \frac{1}{q'^2 + m^2} \>
< \> \exp\left[ \> i q' \cdot( x(t_1) - x(t_2) ) \> \right ]
\> >_{\tilde S_t}
\label{Fourier average S1} \> .
\end{equation}
The path integrals in this expression can be performed analogously
to the equivalent expression for the propagator in Ref.~\cite{RS1}.
Furthermore, the integral over $q'$ may also be done with the help
of the momentum space representation for the propagator
\begin{equation}
\frac{1}{q'^2 + m^2} = \frac{1}{2} \> \int_0^{\infty} du \>
\exp \left [ - \frac{u}{2} \> (q'^2 + m^2) \right ] \>.
\end{equation}
One obtains, after shifting the $u$ integration in the same way as
in  Ref.~\cite{RS1},
\begin{equation}
< \> S_1 \> >_{\tilde S_t} = \> - \frac{g^2}{16 \pi^2} \>
\int_0^{\beta} dt_1 dt_2 \> \frac{1}{\tilde \mu^2(\sigma,T)}
\int_0^1 du \> \> e\left ( m \tilde \mu(\sigma,T),\>
\frac{- i \lambda  W^{(n)}}{ \tilde \mu(\sigma,T)}\> , \> u
\right ) \>.
\label{tilde Fourier average S1 explicit'}
\end{equation}
The quantity $\tilde \mu^2(\sigma, T)$ is defined in terms of the
pseudotime $\mu^2(\sigma,T)$ as in  Ref.~\cite{RS1}, i.e.
\begin{eqnarray}
\tilde \mu^2(\sigma,T) \> &=&\>  \frac{\sigma^2}{A_0 \beta} \>
+ \> \mu^2(\sigma,T) \nonumber \\
 &=&\>  \frac{\sigma^2}{A_0 \beta} \> +\beta \> \sum_{k=1}^{\infty}
\frac{\lambda_k^2}{A_k}.
\label{tilde amu2}
\end{eqnarray}
Here
\begin{equation}
\lambda_k = \frac{\sqrt{2}}{k \pi} \left [ \>
\sin \left(\frac{k \pi t_1}{\beta} \right ) \> - \>
\sin \left(\frac{k \pi t_2}{\beta} \right ) \> \right ] \> .
\label{lambda k}
\end{equation}
and we have introduced the relative and total time
$\sigma = t_1 - t_2$ and $T = (t_1 + t_2)/2$. Finally, the
function $e(s,t,u)$ is defined as
\begin{equation}
e(s,t,u) =
\exp \left ( -  \> \frac{s^2}{2} \> \frac{1-u}{u} \> - \>
\frac{t^2}{2}\> u \> \right ) \> .
\label{e(s,t,u)}
\end{equation}
It should be noted that the interaction term
Eq.~(\ref{tilde Fourier average S1 explicit'}) looks identical to
the interaction term
 written down for the propagator in Ref.~\cite{RS1}, Eq. (97).
Indeed, all the dependence on the number of external mesons is
hidden in the (four dimensional) quantity $W^{(n)}$:
\begin{equation}
W^{(n)}={\sigma \over \beta} P - \sqrt{{\beta \over 2}}
\sum_{k=1}^{\infty} \lambda_k Q_k^{(n)}.
\label{eq:Wn Fourier}
\end{equation}
For $n=0$ this just reduces to $\sigma$ times the nucleon momentum
$p$, while for general $n$ it contains the full complexity
necessary for the $(2 + n)$-point function $G_{2,n}$.
Indeed, using the standard formula for the summation of a series of
cosines weighted by $1/k^2$~\cite{GrRy}, we may evaluate the sums
appearing in $W^{(n)}$ to obtain
\begin{equation}
W^{(n)}=  {\sigma \over 2 } (p_1 + p_{n+1}) + {1 \over 2}
\sum_{i=1}^{n} q_i \> \bigl ( \> \left | \tau_i - t_1 \right | -
\left | \tau_i - t_2 \right | \> \bigr ) \> .
\label{eq:Wn abs}
\end{equation}
Hence $W^{(n)}$ depends on the ordering of the proper times
$\tau_i$ at which the nucleon interacts with the external mesons
relative to the proper times $t_{1,2}$ at which the
internal meson is emitted and absorbed. The form of Eqs. (\ref{eq:Wn Fourier})
and (\ref{eq:Wn abs}) bears some similarity
to the ``phase averaged'' approximation for the Green function and
corrections to it which have been investigated in Ref. \cite{FrGa}.

\begin{figure}
\begin{center}
\begin{picture}(100,70)
\put(25,-15){\makebox(100,70){\psfig{figure=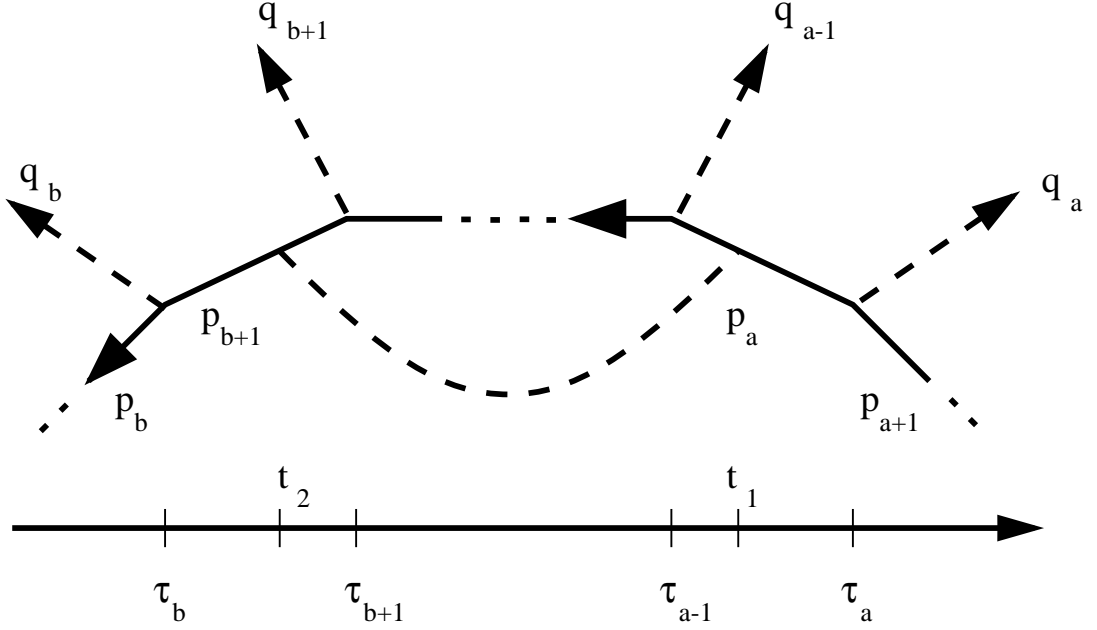}}}
\end{picture}
\end{center}
\caption{The relative ordering of the interaction times of the
external mesons with respect to those of the internal one. It is
assumed that $\sigma$ $>$ 0.}
\label{fig:a and b}
\end{figure}

It is instructive to eliminate the absolute values appearing in
Eq.~(\ref{eq:Wn abs}) by explicitly breaking up the sum
into three contributions: one contribution ($i = 1$ to $b$, say)
where
the internal meson's interaction times $t_{1,2}$ are
both larger than all $\tau_i$, one contribution
where one of them is smaller and the other one is larger than all
$\tau_i$ ($i = b+1$ to $a-1$) and one contribution where they are
both smaller than all $\tau_i$ ($i = a$ to $n$).
This division of the sum is illustrated in Fig.~\ref{fig:a and b}.
Similarly to Fig.~\ref{relative},
Fig.~\ref{fig:a and b} should not be interpreted in the sense of a
Feynman diagram, although it is closely related.
It is a diagrammatic representation of the (in this case averaged)
{\it action} (not an amplitude) in the particle
representation.  It serves to indicate the relative ordering of
the proper times involved. Furthermore, it also defines the momenta:
similarly to a Feynman diagram, four-momentum
is conserved at each vertex, with the proviso that one
associates 0 four-momentum with the
internal meson line (the momentum flowing
though the loop has already been integrated over).

\unitlength1mm
\noindent
One obtains
\begin{eqnarray}
W^{(n)} &=& {\rm sign} (\sigma) \biggl \{\> \bigl [ \tau_{b+1} -
\min(t_1,t_2)\bigr ] p_{b+1} + \sum_{i=b+2}^{a-1} \alpha_i p_i
\nonumber \\
&&\hspace{6cm} + \bigl [ \max(t_1,t_2) - \tau_{a-1} \bigr ] p_{a}
\> \biggr \} \hspace{1cm} a > b + 1 \nonumber \\
&=& \sigma p_{a} \hspace{11cm} a = b + 1 \> .
\label{eq:wn}
\end{eqnarray}
(For the case $a = b + 2 $, the sum appearing in Eq.~(\ref{eq:wn}) is
defined to be empty.) In other words, $W^{(n)}$ is equal to
(up to a sign, which is not relevant as only the
square of $W^{(n)}$ enters) the integral of the proper time
multiplied by the nucleon's four-momentum for the duration of the
exchange of the internal meson. The reason
for the appearance of this quantity will
become clear once we make the connection to perturbation theory.
Particular cases of interest, for the purpose of the eventual
truncation of the Green function, are those where $t_{1,2}$
are both smaller or both larger than all $\tau_i$,  in which case
$W^{(n)}$ $ = $ $\sigma p_1$ and $\sigma p_{n+1}$, respectively.

As was shown in Ref.~\cite{RS1}, if one is interested in the Green
function for the case where the external nucleon legs ({\it not}
the internal ones) are on-shell, then one should consider the
above expression in the limit where the
proper times corresponding to the external legs (i.e. $\alpha_1$ and
$\alpha_{n+1}$, and hence $\beta$) tend towards infinity.  To
implement this limit, the following equality derivable from the
Poisson summation formula~\cite{Light,Act} is
extremely useful: If $F(k)$ is only a function of $k \pi /\beta$ and
if it is even, then up to exponentially small terms in $\beta$
\begin{equation}
\sum_{k=1}^{\infty} \> F( k )\> \simeq \>
\frac{\beta}{\pi} \int_0^{\infty} dE \> F \left ( {E \beta \over \pi}
\right ) - \frac{1}{2} F(0).
\label{Poisson even}
\end{equation}
As was shown in Ref.\cite{RS1}, $A_k$ satisfies these criteria,
which results in the following simplifications:
\begin{equation}
\lim_{\beta \to \infty} \tilde \mu^2(\sigma,T) \> \equiv
\mu^2(\sigma) \>  =
\>  \frac{4}{\pi} \int_0^{\infty} dE \> \frac{1}{A(E)} \>
\frac{\sin^2 (E \sigma/ 2)}{E^2} \>.
\label{amu2(sigma)}
\end{equation}
and
\begin{eqnarray}
2 \> \lim_{\beta \to \infty}\> \sum_{k=0}^{\infty}\> \left [ \>
\log {A_k}\> +\> {1 \over A_k}
\> -\> 1\> \right ]\>
&=& \> \frac{2 \beta }{\pi} \> \int_0^{\infty} dE \>
\left [ \> \log A(E) \> + \> \frac{1}{A(E)} \> - \> 1 \>
\right ] \>\nonumber \\
&&\hspace{1.5cm}  +  \> \log A(0) \> + \> \frac{1}{A(0)} \>
- \> 1 \nonumber \\
&\equiv&  \> \log A(0) \> + \> \frac{1}{A(0)} \> - \> 1 \> + \>
\Omega \> \sum_{i=1}^{n+1\>} \alpha_i \> .
\label{Omega(beta) Poisson}
\end{eqnarray}
Here $\Omega$ is identical to the kinetic term defined in
Ref.~\cite{RS1}.

Using the above simplifications for the case of on-shell external
nucleon legs, as well as the fact that the integrand appearing in
the interaction term $< S_1 >_{\tilde S_t}$ is even
in $\sigma$, one can now write down the expression for the
(untruncated) $(2 + n)$-point function:
\begin{eqnarray}
G_{2,n}(p,p';\{q\})&=& {N_0 \over 2 g}  \sum_{{\cal P}\{q_i\}}
\left \{ \prod_{i=1}^{n+1} g \> \int_0^\infty d\alpha_i \>
\exp \left [ - {\alpha_i \over 2}
\left ( M_0^2 + 2 \Omega + p_i^2 [ 1 - (1 - \lambda)^2 ] \> \right )
\right ] \>  \right \}
\label{eq:unrenormalized}\\
& &\hspace{1cm} \cdot \>
\exp \left [ \frac{g^2}{8 \pi^2} \> \int_0^{\beta} dt_1
\int_{0}^{t_1} dt_2 \> {1 \over \mu^2(\sigma)}
\int_0^1 du \> \> e\left ( m  \mu(\sigma),\>
\frac{- i \lambda  W^{(n)}}{\mu(\sigma)}\> , \> u
\right )\right ] \>.\nonumber
\end{eqnarray}
Here $N_0$ is defined to be
\begin{equation}
N_0 = {1 \over A(0)}\>\exp \left ( 1 - \frac{1}{A(0)} \right )
\end{equation}
 and it should be noted that $G_{2,n}(p,p';\{q\})$ is normalized in
such a way that the correct tree level
result is obtained in the limit in which the coupling vanishes:
\begin{equation}
G_{2,n}^{\rm tree}(p,p';\{q\})\> = \> {1 \over 2 \; g}
\sum_{{\cal P}(\{q\})}  \prod_{i=1}^{n+1} \>
{2\;g \over {p_i}^2 +  M^2_{\rm phys}}
\label{gnfree}
\end{equation}
Furthermore, for non-zero couplings Eq.~(\ref{eq:unrenormalized})
needs to be renormalized in order
for it to become meaningful.  This is because
$\mu^2(\sigma) \to \sigma $ for small $\sigma$ and
hence the integral in Eq.~(\ref{eq:unrenormalized}) diverges
logarithmically. Indeed, the mass renormalization was already
performed in Ref.~\cite{RS1} by demanding that
the pole of the nucleon propagator occurs at the physical nucleon
mass. Actually, the nucleon propagator is just the special case of
the Green function in Eq.~(\ref{eq:unrenormalized}) with $n = 0$.
It is therefore not surprising, but nevertheless crucial for the
internal consistency of the variational approach described here,
that with the same renormalization on the poles  of the external legs
we arrive at the {\it same} relationship between the unrenormalized
nucleon mass $M_0$ and the physical nucleon mass $M$ as one did
in Ref.~\cite{RS1}:
\begin{equation}
M_0^2 + 2 \> \Omega = M^2 [ 1 - (1 - \lambda)^2 ] +
\frac{g^2}{4 \pi^2} \> \int_0^{\infty} { d\sigma \over \mu^2(\sigma)}
\int_0^1 du \> \> e\left ( m  \mu(\sigma),\>
\frac{\lambda  \sigma M}{\mu(\sigma)}\> , \> u \right )
\label{eq:MM0}
\end{equation}
It should be understood, of course, that in principle one should
regularize both Eqs.~(\ref{eq:unrenormalized})
and~(\ref{eq:MM0}). Only after substitution of the latter into the
former does everything become finite
and therefore meaningful with the regulator removed.

\begin{figure}
\begin{center}
\begin{picture}(100,95)
\put(25,-20){\makebox(100,95){\psfig{figure=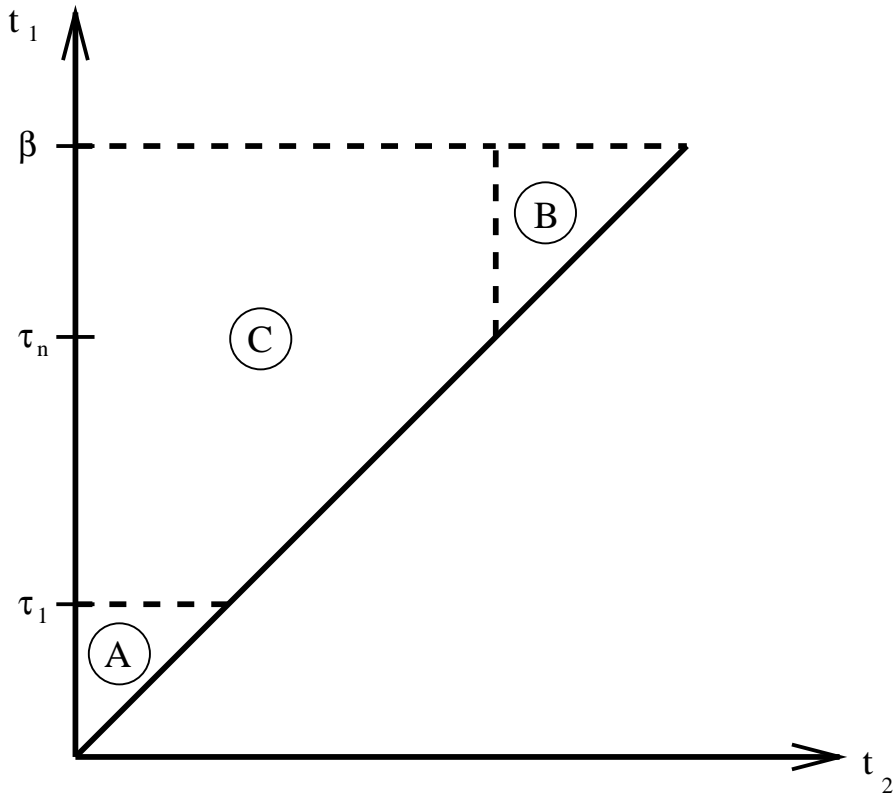}}}
\end{picture}
\end{center}
\caption{The $t_1$ -- $t_2$ integration region. Regions ($A$) and
($B$) correspond to dressing of the external legs and
give rise to the external propagators and residues}
\label{fig:region}
\end{figure}

Because the meaning of $W^{(n)}$ depends on the relative ordering
of the proper times $t_{1,2}$ with respect to the interaction times
$\tau_i$, the $t_1$ -- $t_2$ integration region in
Eq.~(\ref{eq:unrenormalized}) needs to be broken up in order to do
explicit calculations.  In particular, in order to isolate those
parts of the expression which contribute to the dressing
of the external legs one needs to separate out the parts of the
integration region for which $t_{1,2}$ are either both smaller than
all $\tau_i$ (Region ($A$) in Fig.~\ref{fig:region}) or both
larger than all $\tau_i$  (Region ($B$) in Fig.~\ref{fig:region}).
Explicitly, in the limit in which $\alpha_1$ and $\alpha_{n+1}$
become very large, the relevant integrals for these two regions become
\begin{eqnarray}
&&\int_{(A)} dt_1 dt_2 \> \rightarrow \> \int_{0}^{\infty}
d\sigma (\alpha_{1} - \sigma)\nonumber \\
&&\int_{(B)} dt_1 dt_2 \> \rightarrow \> \int_{0}^{\infty}
d\sigma (\alpha_{n+1} - \sigma) \;\;,
\end{eqnarray}
respectively.  Here we have made use of the fact that in regions
($A$) and ($B$) the integrand does not
depend on the overall time $T$.

\unitlength1mm

It is important to note that there is no further dependence on
$\alpha_1$ and $\alpha_{n+1}$ dependence coming
from the integration region $(C)$. This allows one to perform the
integrals over these two proper times.
Comparing with the results for the propagator in Ref.~\cite{RS1},
one sees that the result of these two integrations just yield the
external nucleon propagators, including their residues:
\begin{eqnarray}
&& \int_{0}^{\infty} d\alpha_i \> \exp \Biggl \{ - {\alpha_i \over 2}
\biggl [ ( M^2 + p_i^2 ]\> ( 1 - (1 - \lambda)^2 ) \nonumber \\
& &\hspace{1cm} +  \> \frac{g^2}{4 \pi^2} \> \int_0^{\infty}
 {d\sigma \over \mu^2(\sigma)} \int_0^1 du \> \left (
 e\left ( m  \mu(\sigma),\> \frac{\lambda \sigma M}{\mu(\sigma)}\> ,
 \> u \right ) -
 e\left ( m  \mu(\sigma),\> \frac{ -i \lambda \sigma p_i}
{\mu(\sigma)}\> , \> u \right ) \>
\right ) \>  \biggr ] \> \Biggr \}\nonumber \\
& &\hspace{1cm} \cdot \> \exp \left [
- \frac{g^2}{8 \pi^2} \> \int_0^{\infty}d\sigma
 {\sigma \over \mu^2(\sigma)} \int_0^1 du \>
 e\left ( m  \mu(\sigma),\> \frac{ -i \lambda \sigma p_i}
{\mu(\sigma)}\> , \> u \right ) \>
 \right ] \nonumber\\
&& \hspace{1cm}  = \> {2 Z / N_0 \over p_i^2 + M^2 }\nonumber
\end{eqnarray}
Here $i$ is 1 or $n + 1$, respectively, and $Z$ is the same
expression for the nucleon propagator's residue as
derived in Ref.~\cite{RS2}.

By removing the external legs one may now arrive at the
main result of this paper, the truncated $(2 + n)$-point function
$G_{2,n}^{tr}(p,p';\{q\})$, valid for on-shell external nucleon
momenta:
\begin{eqnarray}
G_{2,n}^{\rm \> tr}(p,p';\{q\})\>&=&\> {2\> g \over N_0} \>
\sum_{{\cal P}\{q_i\}}
\left \{ \> \prod_{i=2}^{n} g \> \int_0^\infty d\alpha_i
\exp \left [ - {\alpha_i \over 2}
\left ( ( M^2 + p_i^2 )( 1 - (1 - \lambda)^2 ) \> \right ) \>
\right ] \> \right \} \nonumber \\
& &\hspace{1.5cm}  \cdot \>
\exp \Biggl \{ - \frac{g^2}{8 \pi^2} \>  \int_0^1 du \>\biggl [
\int_0^{\infty} {d\sigma  \over \mu^2(\sigma)}
\sum_{i=2}^{n}\alpha_i \> e\left ( m  \mu(\sigma),\>
\frac{\lambda \sigma M}{\mu(\sigma)}\> , \> u \right ) \nonumber \\
& &\hspace{2.5cm}
- \> \int_{(C)}   {dt_1 dt_2 \over \mu^2(\sigma)} \>
 e \left ( m  \mu(\sigma),\> \frac{- i \lambda
W^{(n)}}{\mu(\sigma)}\> , \> u \right ) \biggr ] \> \> \>\Biggr \} \>.
\label{eq:truncated}
\end{eqnarray}
In order to appreciate the information contained in
Eq. (\ref{eq:truncated})
it is very instructive
to expand the exponential  in the second and third lines in something
akin to a perturbative expansion (i.e. expand in $g^2$, disregarding
the dependence of $N_0$, $\mu$ and $\lambda$ on $g$).
The leading term contains the tree diagram and for this one the
integrations over the proper times $\alpha_i$ may actually be
carried out, resulting in the appropriate propagators for the
internal nucleon legs. The next term contains
diagrams occuring at one loop order. Because the proper times
$t_1$ and $t_2$ at which the
internal meson couples to the internal nucleon line are integrated
over, they occur in all possible permutations with respect to the
external times $\tau_i$. These permutations give rise to all the
individual Feynman diagrams at this order.
Similar reasoning applies to the higher order terms -- at order
$g^4$ four integrations over the (two) internal mesons' interaction
times must be carried out, each ordering with respect to
the times $\tau_i$ corresponding to a piece contained in a
particular Feynman diagram, etc. Of course,
there is no reason for a particular one of these terms to
reproduce a given Feynman diagram exactly.  This is only guaranteed
for the  one-loop result and is explicitly checked in the Appendix.

Using this procedure, it becomes possible to address the question
whether or not Eq.~(\ref{eq:truncated}) {\it explicitly} contains all
possible Feynman diagrams of any order . (In addition,
it contains additional information {\it implicitly} because, after
all,  $N_0$, $\mu$ and $\lambda$ {\it do} depend on the coupling.
A diagram by diagram identification of this implicit
dependence on the coupling constant does not seem possible.)  In
fact, it is easy to convince oneself that Eq.~(\ref{eq:truncated})
does contain, in an approximate way, almost {\it every} diagram
occurring at {\it any} order in perturbation theory, except for an
important exception. Because of the restriction of the $t_1 - t_2$
integration to Region $(C)$ in Fig. \ref{fig:region},
 diagrams of fourth order or higher  in which $t_1$ and $t_2$ are
both smaller than $\tau_1$ or both larger
than $\tau_n$ (excluding, of course, the one-particle reducible
diagrams taken away in the truncation) are not contained explicitly.
It would be very interesting to see to what extent this feature
depends on the form of the variational function $\rho_k$ in
Eq.~(\ref{general trial action}).

Although Eq.~(\ref{eq:truncated}) is a very compact expression for
all the Green functions in the one nucleon sector of the theory,
it is in general not possible to continue
without specifying the number of external meson legs $n$.  The
simplest case (apart from the nucleon propagator, which has already
been worked out in Refs.~\cite{RS1,RS2}) is the
vertex function, i.e. $n = 1$. We shall discuss this function in
the next Section.  The rather more complicated case of the
$(2 + 2)$-point function (`meson -- nucleon', or `Compton'
scattering) will be discussed in a future paper.

\section{The vertex function}
\label {sec: Vertex}

\noindent
Setting $n = 1$ in the expression for the general truncated Green
function $G_{2,n}^{\rm \> tr}(p,p';\{q\})$ in Eq.~(\ref{eq:truncated})
yields the vertex function.
For this function, the only possible values for $a$ and $b$, as
defined in Fig.~\ref{fig:a and b}, are $2$ and $0$, respectively.
Hence $W^{(n)}$ takes on the form, for positive $\sigma$,
\begin{eqnarray}
W^{(1)} &=& (\alpha_1 - t_2) p_1 + (t_1 - \alpha_1) p_2\nonumber \\
&=&{\sigma \over 2} ( p_1 + p_2 )\> + \>(  T  - \alpha_1 ) q_1\;\;\;.
\end{eqnarray}
It is important to note that although $W^{(1)}$ depends on
$\alpha_1$, the
integral in the last line in Eq.~(\ref{eq:truncated}) does not.  The
$\alpha_1$ dependence disappears by a shift of variable: the
integration region ($C$) for the vertex function is given by, in
the limit of $\alpha_1 \rightarrow \infty$
appropriate for on-shell external nucleons,
\begin{equation}
\int_{(C)} dt_1 dt_2 = \int_{0}^{\infty} d\sigma \>
\int_{\alpha_1 - \sigma/2}^{\alpha_1 + \sigma/2} dT.
\end{equation}
Letting $T \rightarrow T - \alpha_1$ eliminates both the
$\alpha_1$ dependence in the integral limits as well as in $W^{(1)}$.
This is absolutely necessary -- after all, the variable $\alpha_1$
has already been integrated over, resulting in one of the
external propagators of the untruncated Green function.
With this shift, the scattering amplitude is given by
\begin{eqnarray}
A_{2,1}(p,p';q)\>&=&\> {2\> g \> Z \over N_0} \>
\exp \Biggl \{   \frac{g^2}{8 \pi^2} \>  \int_0^1 du
\int_{0}^{\infty}   {d\sigma \over \mu^2(\sigma)} \>
\int_{-\sigma/2}^{\sigma/2} dT
\label{eq:amp1} \\
&&\hspace{2.5cm}
 \cdot \> \exp \left [ -{1 \over 2} \left ( m^2  \mu^2(\sigma)
{1 - u \over u} - {\lambda^2 ({\sigma \over 2}(p_1 + p_2) + T q_1)^2
\over \mu^2(\sigma)} u \right) \> \right ] \> \Biggr \} \> .
\nonumber
\end{eqnarray}
Note the extra factor of $Z$ relating the Green function in
Eq.~(\ref{eq:truncated}) and the corresponding amplitude.

As it stands, the integrand in Eq.~(\ref{eq:amp1}) looks as if it
may diverge for small $\sigma$ due to the factor $1/\mu^2(\sigma)$.
It is easy to see that this is actually
not the case by scaling the $T$ integral by $\sigma$.  That we do not
need an additional (coupling constant) renormalization is, of course,
a special feature of the Wick-Cutkosky model in the quenched approximation
which is a superrenormalizable theory.
By also making
use of the fact that $p_1^2 = p_2^2 = - M^2$, one obtains the
form factor as a function of the momentum transfer $ q^2 = - q_1^2$
in Minkowski space
\begin{eqnarray}
A_{2,1}(q^2)\>&=&\> {2\> g \> Z \over N_0} \>
\exp \Biggl \{  \frac{g^2}{8 \pi^2} \>  \int_0^1 du
\int_{0}^{\infty} d\sigma  {\sigma \over \mu^2(\sigma)} \>
\int_{0}^{1} dt
\label{eq:form factor}\\
&&\hspace{2.5cm} \cdot \>
 \exp \left [ -{1 \over 2} \left ( m^2  \mu^2(\sigma)
{1 - u \over u} + {\sigma^2 \> \lambda^2\over \mu^2(\sigma)}
\left (M^2 - (1 - t^2){q^2 \over 4}  \right ) u \> \right )\right ]
\Biggl \} \nonumber
\end{eqnarray}
Note that, in contrast to perturbation theory, for example, the
change from Euclidean space to Minkowski space was trivial here as
 there are no integrations over momenta to be performed.  Of course,
Eq.~(\ref{eq:form factor}) only defines the
amplitude in the region in which the integrals converge. This is
the case for spacelike momentum transfers, as well as (unphysical)
timelike momentum transfers as long as
 $q^2 < 4 M^2$. In these regions the amplitude is purely real.  For
timelike momentum transfers larger than $4 M^2$, however, one needs
to analytically continue Eq.~(\ref{eq:form factor}) before obtaining
numerical results.  It is possible
to do this if the analytic structure of $\mu^2(\sigma)$ is known.
In particular, if $\mu^2(\sigma)$ has no cuts or poles for
Re$(\sigma) > 0$, then one may deform the contour of the $\sigma$
integration in Eq.~(\ref{eq:form factor})
to the imaginary axis. If one gives $q^2$ a small positive
imaginary part, one must perform the deformation in the positive
half-plane so that one does
not get a contribution from the contour at infinity.
Setting $\sigma = i s$, one obtains
\begin{eqnarray}
A_{2,1}(q^2)\>&=&\> {2\> g \> Z \over N_0} \>
\exp \Biggl \{- \frac{g^2}{8 \pi^2} \>  \int_0^1 du
\int_{0}^{\infty} ds  {s \over \mu^2(i s)} \>
\int_{0}^{1} dt
\label{eq:timelike}\\
&&\hspace{2.5cm} \cdot \>
 \exp \left [ -{1 \over 2} \left ( m^2  \mu^2(i s)
{1 - u \over u} + {s^2 \> \lambda^2\over \mu^2(i s)}
\left ((1 - t^2){q^2 \over 4} - M^2 \right ) u \right) \> \right ]
\Biggr \} \> .
\nonumber
\end{eqnarray}
It is clear that for $q^2 > 4 M^2$ the amplitude is no longer
purely real.  Indeed, it develops a cut, corresponding to the fact
that this is the physical region for
nucleon pair-production. It should of course be remembered that
the calculation has been performed in the quenched approximation;
hence the numerical results in this region certainly do not
correspond to what one would obtain in the full theory.

\section{Numerical Results and Discussion}
\label {sec:Discussion}

\noindent
Our numerical results for the vertex function will be presented as a function
of the dimensionless
coupling constant
\begin{equation}
\alpha = {g^2  \over 4 \pi M^2}\;\;\;.
\label{def alpha}
\end{equation}
We only consider $\alpha$ below the critical coupling $\alpha_c \simeq 0.82$.
For couplings larger than $\alpha_c$
the variational equations do not have real solutions due to the well-known
instability of the Wick-Cutkosky model\cite{RS1,RS2}.

We first investigate the vertex function for low values of the momentum
transfer. Writing Eq.~(\ref{eq:form factor}) as
\begin{equation}
A_{2,1}(q^2) \> = \> A_{2,1}(0) \> F(q^2)
\end{equation}
one may easily read off the effective coupling at zero momentum transfer
and the
elastic form factor for scattering of mesons from the dressed
particle. The effective
coupling takes on a particularly simple form as it may be simplified
through the use of the variational equation for the parameter
$\lambda$ (see Ref.~\cite{RS2}, Sec. 5.1).  One obtains
\begin{equation}
2 g_{{\rm eff}} = A_{2,1}(0) = { 2 g \over \lambda}\;\;\;.
\end{equation}
Like the famous anomalous magnetic moment of the electron in QED
the effective coupling is enhanced because the `velocity parameter'
$\lambda$ is always less than unity. It is interesting to
note that in zeroth variational order we had obtained the enhancement
$g_{{\rm eff}} = g \> A_0 $ \cite{SRA} which can be shown to be less than
the one obtained in first order. Indeed, solving the variational equations
perturbatively for $ m = 0 $ we obtain
\begin{eqnarray}
\frac{g_{\rm eff} - g}{g} \> = \> \> \left \{ \begin{array}{ll}
                          \frac{\alpha}{2 \pi} + . . .   &\hspace{1 cm}
                               \mbox{( zeroth order )} \\
                          \frac{\alpha}{\pi}  + . . .   &\hspace{1 cm}
                                          \mbox{( first order )} \> .
                          \end{array}
                          \right .
\end{eqnarray}
In addition $A_0$ varies much more among the different
parametrizations in Ref.\cite{RS2} of the retardation function than does the
parameter $\lambda$
which is essential for determining the critical coupling.

The elastic form factor is given by
\begin{eqnarray}
F(q^2) &=& \exp \Biggl \{ - \frac{g^2}{8 \pi^2} \int_0^1 du \> \int_0^{\infty}
d\sigma \> \frac{\sigma}{\mu^2(\sigma)} \> \exp \biggl [ -
\frac{1}{2} m^2 \mu^2(\sigma) \frac{1-u}{u} -
\frac{\lambda^2 M^2 \sigma^2}{2 \mu^2(\sigma)} \biggr ] \nonumber \\
&& \hspace{2cm} \cdot \> \int_0^1 dt \> \biggl [ \> 1\> - \>
\exp \biggl ( \frac{\lambda^2 \sigma^2}{8 \mu^2(\sigma)}
(1-t^2) q^2 \> u \> \biggr ) \> \biggr ] \> \Biggr \} .
\label{eq:elastic form factor}
\end{eqnarray}
   From its low-$q$ expansion we deduce
the mean square radius of the dressed particle to be
\begin{equation}
\left < r^2 \right >  =  \frac{g^2}{16 \pi^2} \> \lambda^2 \int_0^1 du \>u
\int_{0}^{\infty} d\sigma  {\sigma^3 \over \mu^4(\sigma)} \>
 \exp \left [ -{1 \over 2} \left ( m^2  \mu^2(\sigma)
{1 - u \over u} +
{ \lambda^2 M^2 \sigma^2 \over \mu^2(\sigma)}
 u \right) \> \right ] \;\;\;.
\label{eq:radius}
\end{equation}
In the limit $\lambda \to 1, \mu^2(\sigma) \to \sigma $ we can perform the
$\sigma$-integration and recover the perturbative result
\begin{equation}
 \left < r^2 \right >_{\rm pert} \> = \> \frac{g^2}{4 \pi^2} \>
\int_0^1 du \> \frac{u^3}
{ \left [ \> (1-u) m^2 + u^2 M^2 \> \right ]^2 } \> \> .
\label{pertubative radius}
\end{equation}
For large spacelike momenta $q^2 \ll - M^2$ the elastic form factor
(\ref{eq:elastic form factor}) approaches the constant value
\begin{equation}
F(q^2) \> \longrightarrow \> \exp \Biggl \{ - \frac{g^2}{8 \pi^2}
\int_0^1 du \> \int_0^{\infty}
d\sigma \> \frac{\sigma}{\mu^2(\sigma)} \>
 \exp \left [ -{1 \over 2} \left ( m^2  \mu^2(\sigma)
{1 - u \over u} +
{ \lambda^2 M^2 \sigma^2 \over \mu^2(\sigma)}
 u \right) \> \right ] \>  \Biggr \}
\end{equation}
which is identical with the constant $ N_1 = \lambda Z / N_0 $
encountered in
the calculation of the residue \cite{RS2}. Although the large $q$-limit
certainly is outside the region of applicability of the quenched model
the result still is meaningful: for asymptotically large momentum
transfers
the meson ``sees'' the bare particle inside the Wick-Cutkosky polaron.

The effective coupling constant and the radius of the dressed particle
are plotted as a function of the dimensionless coupling
in Figs.~\ref{fig:coupling} and~\ref{fig:radius}. We have employed
several parametrizations of the retardation function
in the quadratic trial action
\footnote{See Refs.~\cite{RS1,RS2} for more details about these
trial actions;
 the `Feynman' trial action is of the same form as that used in the
variational calculation for the polaron~\cite{Fey1} while the
`improved' trial action has the same ultraviolet
behavior as the true action. The `extended' parametrization gives in
addition the correct threshold behaviour in zeroth order meson-nucleon
 scattering
\cite{SRA} and leads to the lowest value of the variational functional
among the three parametrizations.}.  The
nucleon mass is taken to be 939 MeV, while the meson mass is fixed
at 140 MeV. For comparison, the perturbative and zeroth order results are
also shown.
Table 1 demonstrates quantitatively
that the results of the variational calculation are practically
independent
of which ansatz for the trial action is chosen.
Indeed, if one leaves the form
of the profile function free and lets the variational principle
determine it (corresponding to the `variational' trial action in
Ref.~\cite{RS1}), the results for the
effective coupling and radius are nearly indistinguishable from those
of the `improved' or the `extended' parametrization.
Furthermore, the results agree with
perturbation theory for small couplings while they
deviate quite strongly for large couplings.

\begin{table}
\begin{center}
\begin{tabular}{|c|c|c|c|c|c|} \hline
{}~~$\alpha$~~ & ~`Feynman'~ & ~`improved'~   &~`extended'~
& ~`variational'~ & ~perturbative~ \\ \hline
 0.1   & 0.0480   & 0.0481   & 0.0481      & 0.0481   & 0.0470 \\
 0.2   & 0.0697   & 0.0697   & 0.0697      & 0.0697   & 0.0664 \\
 0.3   & 0.0878   & 0.0879   & 0.0879      & 0.0879   & 0.0813 \\
 0.4   & 0.1047   & 0.1049   & 0.1049      & 0.1049   & 0.0939 \\
 0.5   & 0.1216   & 0.1220   & 0.1220      & 0.1220   & 0.1050 \\
 0.6   & 0.1397   & 0.1404   & 0.1404      & 0.1404   & 0.1150 \\
 0.7   & 0.1610   & 0.1622   & 0.1622      & 0.1623   & 0.1242 \\
 0.8   & 0.1937   & 0.1980   & 0.1979      & 0.1985   & 0.1328 \\
\hline
\end{tabular}
\end{center}
\caption{
First-order root-mean-square radius (in fm) of the dressed particle
from Eq. (\protect\ref{eq:radius}) for various parametrizations
of the retardation function. The heading `Feynman'
gives the result in the Feynman parametrization whereas
`improved' refers to the `improved' parametrization. In both cases
the parameters have been determined from
the variational calculation for the nucleon self-energy in
Ref.~(\protect\cite{RS2}) . The `extended' parametrization was introduced in
Ref.~(\protect\cite{SRA})
for additional use in meson-nucleon scattering.
The radius calculated with the solution of
the variational equations is denoted by
`variational'. For comparison the perturbative result is also given.}
\end{table}

\unitlength1mm
\begin{figure}
\begin{center}
\begin{picture}(80,120)
\put(170,160){\makebox(80,80){\psfig{figure=fig4a.ps}}}
\put(170,75){\makebox(80,80){\psfig{figure=fig4b.ps}}}
\end{picture}
\end{center}
\caption{The effective coupling strength as a function of the
dimensionless coupling  $\alpha$ : (a) in zeroth order, (b) in first order of
the variational calculation.
The solid (dashed) line corresponds to the use of the `improved'
(`Feynman') trial actions, while
the dash-dotted line is the result from the `extended' parametrization.
The dotted line denotes the perturbative result.}
\label{fig:coupling}
\end{figure}

\unitlength1mm
\begin{figure}
\begin{center}
\begin{picture}(80,120)
\put(170,160){\makebox(80,80){\psfig{figure=fig5a.ps}}}
\put(170,75){\makebox(80,80){\psfig{figure=fig5b.ps}}}
\end{picture}
\end{center}
\caption{The mean square radius as a function of the dimensionless
coupling  $\alpha$ : (a) in zeroth order, (b) in first order of
the variational calculation.
The various curves have the same meaning as
those in Fig.~\protect\ref{fig:coupling}.}
\label{fig:radius}
\end{figure}

\begin{figure}
\begin{center}
\begin{picture}(80,170)
\put(0,120){\makebox(80,80){\psfig{figure=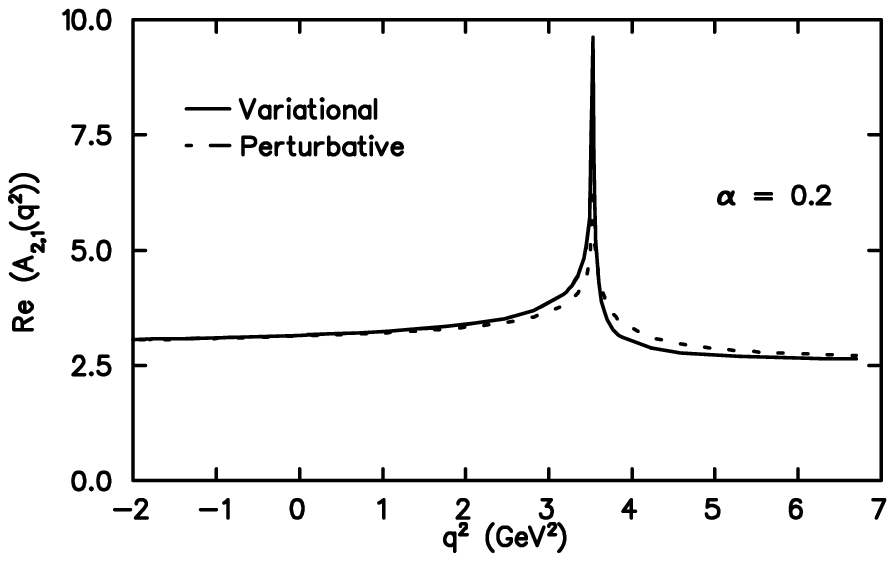}}}
\put(0,55){\makebox(80,80){\psfig{figure=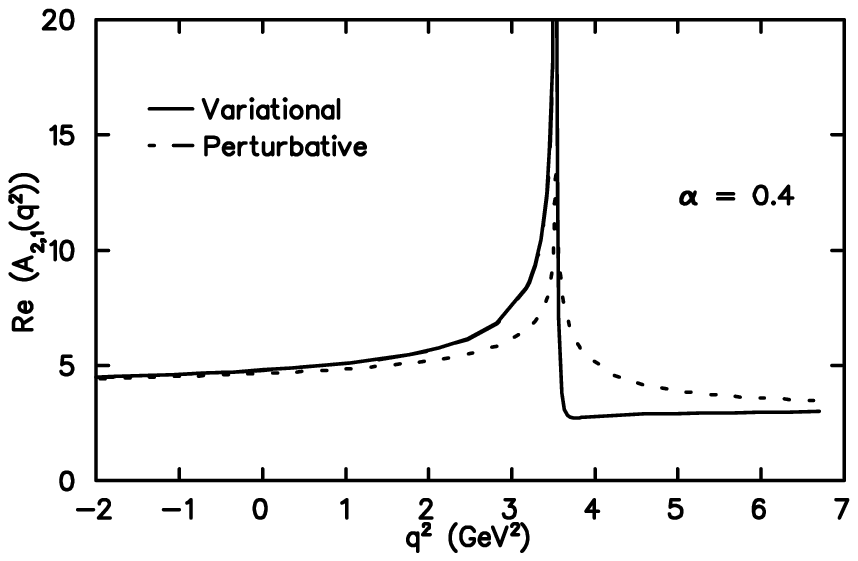}}}
\put(0,-10){\makebox(80,80){\psfig{figure=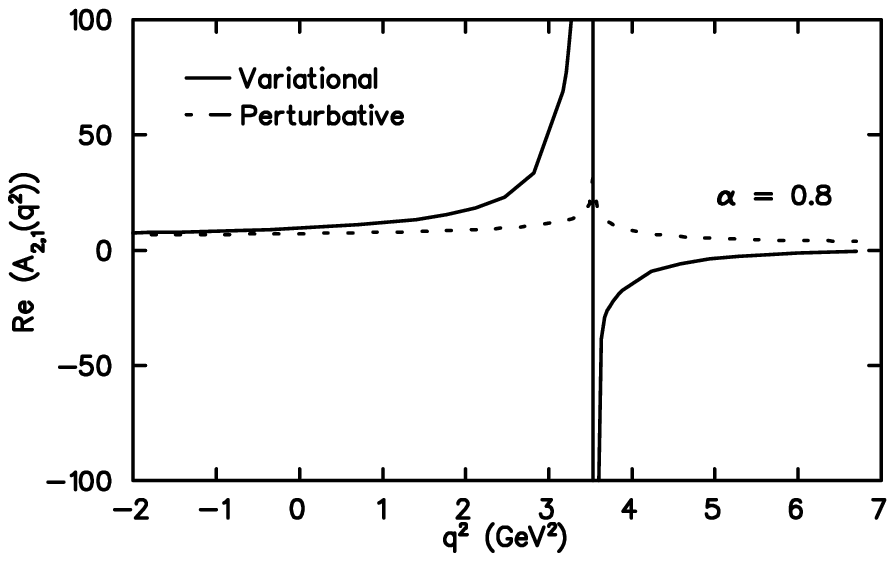}}}
\end{picture}
\end{center}
\caption{The real part of the amplitude as a function of the
momentum transfer squared, for
from top to bottom, $\alpha$ = 0.2 (a), 0.4 (b) and 0.8 (c)  The
solid  line corresponds to the use of the `improved'  trial action,
while the dotted line is the perturbative result}
\label{fig:amp_real}
\end{figure}

Even though there are large numerical differences between the
perturbative and variational calculation of the effective coupling
and radius, one does not observe a
qualitative difference.  The situation is dramatically different
for the form factor away from zero momentum transfer.
The real part is plotted, for three
different couplings, as a function of the
meson's momentum transfer squared in Fig.~\ref{fig:amp_real}.
For $q^2 \leq 4 M^2$  the representation
of Eq.~(\ref{eq:form factor}) was used, while for $q^2 \geq 4 M^2$ we
used the analytically continued representation of
Eq.~(\ref{eq:timelike}). The physical region
for meson absorption is $q^2 \leq 0$, while
$q^2 \geq 4 M^2$ corresponds to pair-production. (The unphysical
region $0 < q^2 < 4 M^2$ is also shown because the matching of the
two representations at $q^2 = 4 M^2$ provides a useful
check on the accuracy of the numerical integrations.)

\unitlength1mm

In general, the perturbative and variational calculations tend to
approach each other at large values of $\left | q^2 \right | $.
In the area just above threshold, however,
major differences arise as the coupling is increased.  Whereas the
perturbative amplitude increases smoothly
in magnitude as the coupling is increased from $\alpha = 0.2 $ to
$\alpha = 0.8 $, the variational amplitude  decreases from
$\alpha = 0.2 $ to $\alpha = 0.4 $ and
actually becomes negative for $\alpha = 0.8 $.  As the amplitude
has to be positive at threshold ( Eq.~(\ref{eq:form factor})) is
always
positive) this necessitates a rapid oscillation very close to
threshold.  Furthermore, near threshold the variational
amplitude is numerically greatly enhanced (at $\alpha = 0.8 $ by
orders of magnitude!) as compared to the perturbative result.

\begin{figure}
\begin{center}
\begin{picture}(80,170)
\put(0,120){\makebox(80,80){\psfig{figure=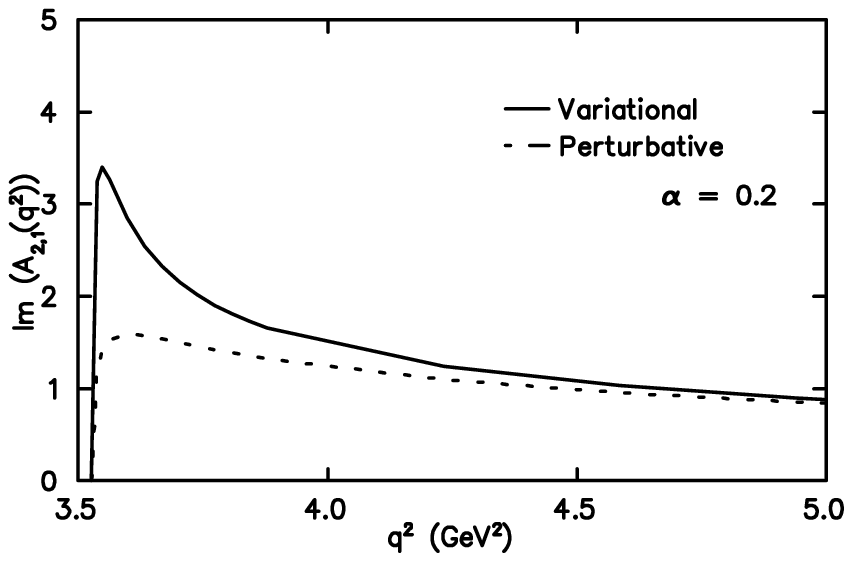}}}
\put(0,55){\makebox(80,80){\psfig{figure=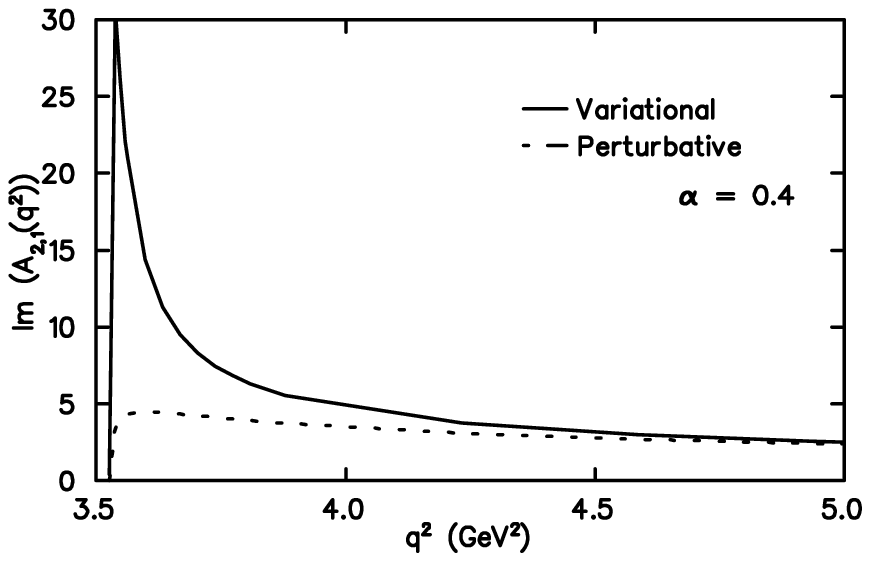}}}
\put(0,-10){\makebox(80,80){\psfig{figure=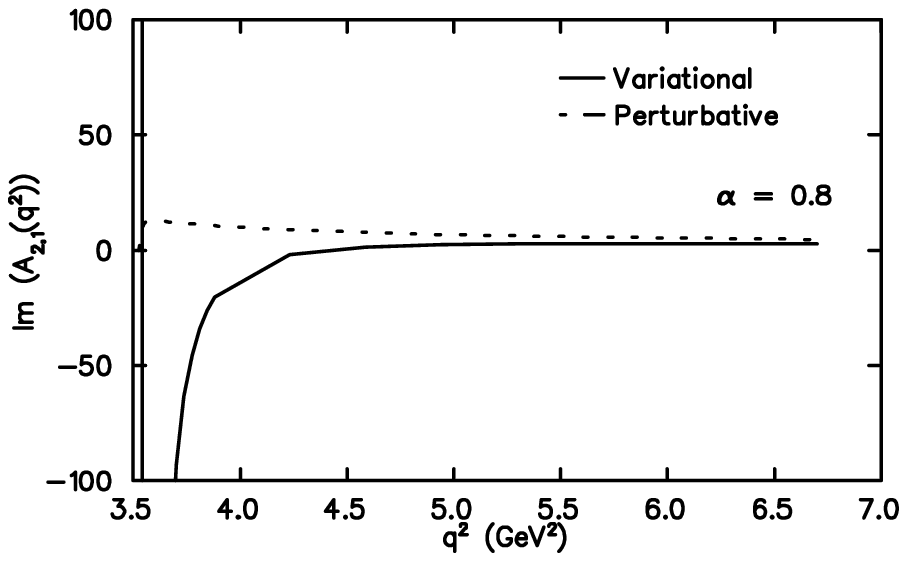}}}
\end{picture}
\end{center}
\caption{The imaginary part of the amplitude as a function of the
momentum transfer squared, for
from top to bottom, $\alpha$ = 0.2 (a), 0.4 (b) and 0.8 (c)  The
solid  line corresponds to the use of the `improved'  trial action,
while the dotted line is the perturbative result}
\label{fig:amp_imag}
\end{figure}

A somewhat similar behavior is exhibited by the imaginary part of
the amplitude, shown in Fig.~\ref{fig:amp_imag}.  This is, of course,
only non-zero above threshold.  Again
the variational calculation and perturbation theory approach
each other for large $q^2$ and the two calculations are
qualitatively similar for the relatively small coupling
of $\alpha = 0.2$ ( Numerically, though, there is already a
factor of two between the two calculations at the peak.)
In contrast to the real part of the amplitude, the imaginary
part of the variational amplitude is still increasing at modest
couplings ($\alpha = 0.4$),
but again at $\alpha = 0.8$ large oscillations show up.

\unitlength1mm

The pronounced quantitative and qualitative differences between
the leading order perturbative calculation and the variational one
are due to the fact that the latter has not been expanded in powers
of the coupling constant. Let us writes  Eq.~(\ref{eq:timelike}) as
\begin{equation}
A_{2,1}(q^2)\>=\> {\rm const.} \>
\exp \left \{\frac{\alpha}{2 \pi} \left [ {\rm Re}\>
\Xi (q^2,\alpha) + i \> {\rm Im}\>
 \Xi (q^2,\alpha)\right ] \right \} \;\;\;,
\label{eq:var}
\end{equation}
where the dependence of $ \> \> \Xi (q^2,\alpha) $ on the coupling arises
from the dependence of the variational parameters on the coupling.
The perturbative amplitude up to one-loop order is just given by
\begin{equation}
A_{2,1}^{pert.}(q^2)\>=\> {\rm const.'} \>\left \{ 1 \> + \>
\frac{\alpha}{2 \pi} \left [
 {\rm Re}\> \Xi (q^2,0) + i \> {\rm Im}\> \Xi (q^2,0)\right ]
\right \} \;\;\;.
\label{eq:pert}
\end{equation}
In the function $ \> \Xi (q^2,\alpha) \> $, as opposed to the complete
amplitude, there are in fact only quantitative, but not qualitative,
differences between the perturbative and the variational
calculations.  For example, the imaginary parts of
$ \> {\rm Im} \> \Xi (q^2,0) \> $ and  $ \> {\rm Im}\> \Xi (q^2,\alpha
= 0.2) \> $
both rise rapidly from 0 at threshold
to less than 20 just above threshold and
then slowly decrease.  At $ \alpha = 0.8 $ the same behavior takes
place, except this time the maximum of the variational calculation
is a little bit more than twice as large.
The origin of the qualitative differences between
Eqs.~(\ref{eq:var})~and~(\ref{eq:pert}) are due to the fact that even
though the coefficient  $ \> \alpha/2 \pi \> $ is not particularly
large (for example, it is only 0.11 for $ \alpha = 0.8 $) -- so one
would expect leading order perturbation theory to
be quite good --  the numerical value of the integral
$ \> \Xi (q^2,\alpha)$ more than compensates for it and makes the
perturbative result meaningless.  In particular,
it is clear from Eq.~(\ref{eq:var}) that the real (imaginary) parts
of the amplitude will have zeros when
$ \> \frac{\alpha}{2 \pi} {\rm Im} \Xi (q^2,\alpha) \> $ is
$\pi / 2$, $3 \pi / 2$, ... (0, $\pi$, ...) while the perturbative
result in Eq.~(\ref{eq:pert}) will not.
Keeping higher terms in the perturbative expansion may in fact also
give rise to zeros, but it is clear that, with
$ \> \frac{\alpha}{2 \pi} {\rm Im}\>  \Xi (q^2,\alpha) \> $ numerically
approaching 5  at the peak for $\alpha = 0.8$, to obtain any
reasonable numerical agreement between the perturbative and the
variational results one would
have to expand the perturbation theory to a very high order indeed.

\section{Conclusion and Outlook}
\label{sec:Conclusion}

\noindent
The main result of this paper is a compact expression for the
truncated Green function of an on-shell scalar nucleon interacting
with an arbitrary number of mesons (Eq.~(\ref{eq:truncated})),
calculated at lowest non-trivial order in the variational approach of
Ref.~\cite{RS1}.  A welcome feature of this non-perturbative approach
to the relativistic field theoretic strong coupling problem is that
the renormalization and the truncation of this general Green function
may be done analytically and in a self consistent manner.  Furthermore,
by construction the expansion of the Green function in the coupling
constant agrees exactly with the result of renormalized 1-loop
perturbation theory. Clearly, this assures that the small coupling
limit of the calculation is correct {\it numerically}.  This was an important
characteristic in the original application of this
variational technique -- numerical agreement with {\it both} the weak
and strong coupling limits was the reason why the famous application
to the polaron problem was so successful.  Furthermore, in the present
case it has the additional consequence that the {\it analytical}
structure of variational Green functions is at least as rich as that
of 1-loop perturbation theory:  cuts due to particle production
thresholds opening up as well as poles due to particle exchange in
the intermediate state are both contained in $G_{2,n}$.
Also, of course, the technique is fully covariant; the Green functions
may be used for any process, be it scattering or particle production, and
indeed are even valid outside the physical regions.

Dependence on the coupling constant of the theory enters the Green
function both explicitly, as well as implicitly through the
variational parameters.  The explicit dependence allows one to
identify contributions to individual Feynman diagrams.  Indeed, one
finds that the expression for the Green function sums up, in an
approximate way, Feynman diagrams up to infinite order in perturbation
theory (the untruncated Green function in fact contains parts of all
possible Feynman diagrams).

Most of the features mentioned above are independent of the functional
form of the trial action.  The analytic structure of the Green
function $G_{2,n}$, however, does seem to depend on this functional
form.  For the present we have been satisfied with
a particular (but still very general) choice for the trial action
(Eqs. (\ref{general trial action}) and (\ref{eq:choice}))
which does allow the non-trivial truncation of the external
dressed nucleon propagators to be made.
Indeed, it results in the {\it same} variational equations which have
already been found in Refs. \cite{RS1,RS2} for the propagator,
independent of the number of external mesons. This has the great
practical advantage in that it means that the determination of the
variational parameters only has to be done once.  It would be
interesting to see if this is a general feature or whether it is
specific to the class of trial actions considered here.

In the second part of this paper we investigated the case of meson
absorption or emission and nucleon pair-production, i.e.  $n = 1$,
numerically.  The much richer case of particle scattering, $n = 2$, is
considered in a following paper.  After a trivial rotation to
Minkowski space, the general expression for $G_{2,n}$ for $n=1$
immediately yields the amplitude in the physical region for meson
absorption/emission (i.e.  $q^2 < 0$) as well as in the unphysical
region $0 < q^2 < 4 M^2$. The representation is not defined for $q^2 >
4 M^2$, but may be analytically continued into this region.  One finds
a cut starting at the pair production threshold.  At this point it
should be remembered that we are working in the quenched
approximation for the nucleons, so the numerical results in this
region will differ to the unqenched ones.  Comparison to other quenched
calculations, such as perturbation theory, do of course remain meaningful.

We have numerically investigated the vertex function as a function of
the momentum transfer and the strength of the coupling.  A strong
coupling limit of the Wick-Cutkosky model does not exist because of
its instability, so a numerical appraisal of the variational method
for very large $\alpha$ has to await its application to a different
field theory. Nevertheless certain non-perturbative properties can already
be addressed in the range of coupling constants available in the
variational treatment of the Wick-Cutkosky model. Among these are the
critical coupling and the width of scalar nucleon for even larger
coupling constants which have been extracted from
the propagator in Ref. \cite{RS2}. The vertex function evaluated in the
present paper provides another example of large nonperturbative
effects: even
for the relatively `small' couplings allowed
in the Wick-Cutkosky model the results differ markedly from what one
would obtain in perturbation theory, both quantitatively and
qualitatively.  The origin of this disagreement may be traced back to
the fact that although the coupling is indeed relatively small,
the integrals
multiplying the coupling are in fact very large, in particular near
threshold -- both in perturbation theory as well as in the variational
calculation.  The lesson to be learned, well known of course, is that
smallness of the coupling is not a good criterion to determine whether
or not to trust a perturbative result.  In the present case, (1-loop)
perturbation theory is inadequate even for values of the dimensionless
coupling constant as small as 0.4 .

In conclusion, we think that the variational method described here is
a rather promising technique for systematically calculating Green functions
in problems where perturbation theory is inadequate. We have shown
that it is possible to truncate Green functions with one nucleon
and an arbitrary number of external mesons in a manner
which is consistent with the variational ansatz.
It remains an interesting problem to extend this treatment to several
external nucleons and to study bound-state problems.
In addition, it seems worthwhile to find out how this approach
will fare in realistic field theories.  Attempts to apply it in QED are
currently under way.

\vspace{2cm}
\noindent {\bf Acknowlegements}

We would like to thank Dina Alexandrou for many fruitful discussions
related to this work and a critical reading of the manuscript.
One of the authors (R.R.) is grateful
to Avraham Rinat for the kind hospitality offered to him at
the Weizmann Institute of Sciences where part of this work was
conceived.

\newpage

\noindent
{\Large\bf Appendix : Perturbative Limit}
\renewcommand{\theequation}{A.\arabic{equation}}
\setcounter{equation}{0}

\vspace{1.5cm}
\noindent
It is a straightforward, but nevertheless useful, exercise to
verify that the truncated $(2 + n)$-point function in
Eq.~(\ref{eq:truncated}) has the correct perturbative limit.  We shall
sketch the derivation of the perturbative result in this appendix.
It suffices to consider a single perturbative diagram
$G_{2,n}^{a,b}(p,p';\{q_i\})$ of the form shown in
Fig.~\ref{fig:a and b}. The complete amplitude is then obtained by a
summation over all a and b as well as a summation over all
permutations of the external meson momenta $q_i$.  To start off with,
let us consider the case where $a >  b + 1$, i.e. a diagram which
has at least one external meson coupling to the nucleon while the
internal meson is being exchanged.
For this type of diagram one may replace, at this order, the
unrenormalized nucleon mass $M_0$ by  the physical mass $M$.
The perturbative truncated
Green function corresponding to this situation is  given by
\begin{eqnarray}
G_{2,n}^{a,b}(p,p';\{q_i\}) &= & {1 \over 2 g}
\left [\prod_{i=2}^{b+1} {2 g \over p_i^2 + M^2}\right ]
\left [\prod_{i=a}^{n} {2 g \over p_i^2 + M^2}\right ]\nonumber \\
& & \hspace{1cm} \cdot \> (2 g)^2 \int {d^4k \over (2 \pi)^4}
{1 \over k^2 +
m^2} \left [\prod_{i=b+1}^{a} {2 g \over (p_i-k)^2 + M^2}\right ]
\end{eqnarray}
It is customary at this stage to rewrite the denominator in terms
of integrals involving Feynman parameters.  For the purposes of
comparison to the variational calculation this is, however, not
very useful as the variational
calculation gives rise to an expression in terms of integrals over
proper time.  Although in general there will be a relation between
these proper times and the Feynman parameters, it will be non-trivial.

Rather, it is easier to proceed by exponentiating all denominators
in the usual way:
\begin{equation}
{1 \over p^2 + M^2} = {1 \over 2}\int_0^{\infty} dx \>
e^{-x(p^2 + M^2)/2 }
\end{equation}
and then performing the gaussian integration over the internal
meson momentum $k$.  If one exponentiates all propagators in this
way one obtains a product
of $n + 2$ integrals, the same number which one obtains if one
expands the exponential appearing in
the truncated Green's function in Eq.~(\ref{eq:truncated}).  Indeed,
 the $n + 2$ integration parameters of these two expressions may be
identified individually if one chooses them in the following way:
A nucleon propagator involving $p_i^2$ or $(p_i-k)^2$ between
{\it external} mesons should just be exponentiated with the
parameter $\alpha_i$.
The two pairs of nucleon propagators between an external meson and
the {\it internal} meson need to be exponentiated a little bit
differently - one needs to recombine the integration parameters in
the following way:
\begin{eqnarray}
{2 \over p_a^2 + M^2}\>{2 \over (p_a-k)^2 + M^2} &=&
\int_0^{\infty}dx\> dy\> \exp \left [ -{x \over 2}(p_a^2 + M^2)
\right ] \> \exp \left [ - {y \over 2}((p_a-k)^2 + M^2) \right ]
\nonumber \\
&=& \int_0^{\infty}d\alpha_a \int_{\tau_{a-1}}^{\tau_a}dt_1 \>
\exp \left [ -{t_1-\tau_a \over 2} ( (p_a-k)^2 - p_a^2) \right ]
\nonumber \\
&& \hspace{2cm} \cdot
\> \exp \left [ - {\alpha_a \over 2}((p_a-k)^2 + M^2) \right ]
\end{eqnarray}
and similarly for the two propagators involving $p_{b+1}$.
Here we have used the transformations $\alpha_a = x + y$ and
$t_1 = \tau_a - x$, the time $\tau_a$ being defined to be the same
sum of $\alpha_i's$ as used in the
main text. The reason for this particular transformation should be
clear from Fig.~\ref{fig:a and b}.  Finally, the integration
parameter associated with
the internal meson should be transformed in the same way as the
parameter $u$ in the main text.

Using these definitions, one readily obtains
\begin{eqnarray}
G_{2,n}^{(a,b)}(p,p';\{q\})\>&=&\> 2\> g
\left \{ \prod_{i=2}^{n} g \> \int_0^\infty d\alpha_i
\> \exp \left [ - {\alpha_i \over 2} ( M^2 + p_i^2 )
\right ] \> \right \} \nonumber \\
& &\hspace{1cm}
\cdot \>\frac{g^2}{8 \pi^2} \> \int_{\tau_{a-1}}^{\tau_{a}} dt_1
\int_{\tau_{b}}^{\tau_{b+1}} dt_2 \int_0^1 {du \over \sigma} \>
 e \left ( m  \sqrt \sigma ,\> \frac{- i \lambda  W^{(n)}}
{\sqrt \sigma }\> , \> u \right ) \>.
\label{eq:pert1}
\end{eqnarray}

To complete the perturbative calculation of the $(n + 2)$-point
function,
 we still need to consider the trivial case where $a = b+1$, i.e.
the class of diagrams where the internal meson loop just
dresses an internal nucleon propagator.  This is the only type of
diagram which diverges and hence needs to be renormalized.
For this class of diagrams one obtains, together with the
tree level term and using the same type of exponentiation as
in the derivation of Eq.~(\ref{eq:pert1}), that

\begin{eqnarray}
G_{2,n}^{(a)}(p,p';\{q\})\>&=&\> 2\> g
\left \{ \prod_{i=2}^{n} g \> \int_0^\infty d\alpha_i  \>
\exp \left [ - {\alpha_i \over 2}  (M^2 + p_i^2 )
\right ] \> \right \} \> \cdot \> \Biggl \{ \> 1   \nonumber \\
& &\hspace{5mm}
- \>\frac{g^2}{8 \pi^2} \> \alpha_a \>\int_{0}^{\infty}
{d\sigma \over \sigma} \int_0^1 du  \>
 e \left ( m  \sqrt \sigma ,\> M \sqrt \sigma , \> u \right )
\nonumber \\
& &\hspace{10mm}
 + \>\frac{g^2}{8 \pi^2} \> \>\int_{\tau_{a-1}}^{\tau_{a}}
{dt_1 dt_2 \over \sigma} \int_0^1 du  \>
 e \left ( m  \sqrt \sigma ,\>{- i W^{(n)} \over  \sqrt \sigma} ,
\> u \right ) \> \Biggr \} \>.
\label{eq:pert2}
\end{eqnarray}
Finally then, we obtain the complete perturbative result for the
truncated $(2 + n)$-point function by summing over all
permutations of the external momenta as well as over $a$ and $b$:
\begin{eqnarray}
G_{2,n}^{\rm \> pert}(p,p';\{q\})\>&=&
\sum_{a=2}^{n} G_{2,n}^{(a)}(p,p';\{q\})\>\>
+ \>\>\sum_{b=0}^{n-1}\sum_{a=b+2}^{n+1} G_{2,n}^{(a,b)}(p,p';\{q\})
\nonumber \\
&=&\> 2\> g
\left \{ \prod_{i=2}^{n} g \> \int_0^\infty d\alpha_i
\exp \left [ - {\alpha_i \over 2}  ( M^2 + p_i^2 )
\right ] \> \right \} \> \cdot \> \Biggl \{ \> 1  \nonumber \\
& &\hspace{5mm}
-  \>\frac{g^2}{8 \pi^2} \int_{0}^{\infty}
{d\sigma \over \sigma}
 \int_0^1 du  \> \sum_{i=2}^{n}\alpha_i \>
 e \left ( m  \sqrt \sigma ,\> M \sqrt \sigma , \> u \right )
\label{eq:pertcompl}\\
& &\hspace{10mm}
 + \>\frac{g^2}{8 \pi^2} \> \>\int_{\tau_{1}}^{\tau_{n}}
{dt_1 dt_2 \over \sigma} \int_0^1 du  \>
 e \left ( m  \sqrt \sigma ,\>{- i W^{(n)} \over  \sqrt \sigma} ,
\> u \right ) \nonumber \\
&& \hspace{15mm}  + \> \frac{g^2}{8 \pi^2} \> \int_{0}^{\beta}
{d\sigma \over \sigma}
\int_{\tau_{1}-\sigma/2}^{\tau_{n}+\sigma/2} dT
 \int_0^1 du  \>
 e \left ( m  \sqrt \sigma ,\> \frac{- i   W^{(n)}}{\sqrt \sigma }
\> , \> u \right ) \> \Biggr \}
 \>.\nonumber
\end{eqnarray}

Remembering that $A(E)$ and  $\lambda$ have perturbative expansions
of the form $1 \> + \> {\cal O}(g^2)$
(and hence $\mu^2(\sigma)$ = $\sigma$ + $ {\cal O}(g^2)$),
it is easily checked that this agrees with the corresponding result
of the variational calculation given in Eq.~(\ref{eq:truncated}).


\begin{thebibliography}{99}

\bibitem{Sch} L. I. Schiff, Phys. Rev. {\bf 130} (1963) 458.

\bibitem{CJT} J. M. Cornwall, R. Jackiw and E. Tomboulis,
Phys. Rev. {\bf D 10} (1974), 2428.

\bibitem{Stev} P. M. Stevenson, Phys. Rev. {\bf D 30} (1984) 1712,
{\bf D 32} (1985) 1389; T.~Barnes and G.~I.~Ghandour, Phys. Rev.
{\bf D 22} (1980) 924.

\bibitem{Fey2} Proc. of the Intern. Workshop on
Variational Calculations in Quantum Field Theory, L. Polley and
D. E. L. Pottinger, Eds., World Scientific (1988).

\bibitem{Hay} R. W. Haymaker, Riv. Nuovo Cim. {\bf 14} (1991),
 no. 8 , 1.

\bibitem{Fey1} R. P. Feynman, Phys. Rev. {\bf 97} (1955), 660.

\bibitem{FeHi} R. P. Feynman and A. R. Hibbs : Quantum Mechanics
and Path Integrals, McGraw-Hill (1965).

\bibitem{RoFe} C. Rodriguez and V. K. Fedyanin,
Sov. J. Part. Nucl. {\bf 15} (1984), 390.

\bibitem{MCM} T. K. Mitra, A. Chatterjee and S. Mukhopadhyay,
Phys. Rep. {\bf 153} (1987), 91.

\bibitem{BoPl} N. N. Bogoliubov Jr. and V. N. Plechko,
Riv. Nuovo Cim.{\bf 11} (1988), 1.

\bibitem{GeLo} B. Gerlach and H. L\"owen, Rev. Mod. Phys.
{\bf 63} (1991), 63.

\bibitem{AlRo} C. Alexandrou and R. Rosenfelder, Phys. Rep.
{\bf 215} (1992), 1.

\bibitem{Mano} K. Mano, Prog. Theor. Phys. {\bf 14} (1955) 435

\bibitem{RS1} R. Rosenfelder and A. W. Schreiber, "Polaron
Variational Methods in the Particle Representation of Field Theory:
I. General Formalism", nucl-th/9504002, to be published.

\bibitem{RS2} R. Rosenfelder and A. W. Schreiber, "Polaron
Variational Methods in the Particle Representation of Field Theory:
II. Numerical Results for the Propagator", nucl-th/9504005,
to be published. See also preprint PSI-PR-94-07.

\bibitem{SRA} A. W. Schreiber, R. Rosenfelder and C. Alexandrou,
"Variational calculation of relativistic meson-nucleon scattering
in zeroth order", preprint PSI-PR-95-08, nucl-th/9504023,
to be published.

\bibitem{Pol1} R. Rosenfelder J. Phys. {\bf A27} (1994) 3523.

\bibitem{Pol2} C. Alexandrou, Y. Lu and  R. Rosenfelder,
Z. Phys. {\bf A 350} (1995) 131.

\bibitem{AlDi} C. Alexandrou and F. K. Diakonos, preprint
UCY-PHY-94/9, nucl-th/9503010, to appear in Z. Phys. {\bf A}.

\bibitem{Wick} G. C. Wick, Phys. Rev. {\bf 96} (1954), 1124.

\bibitem{Cut} R. E. Cutkosky, Phys. Rev. {\bf 96} (1954), 1135.

\bibitem{SiTj} Yu. A. Simonov and J. A. Tjon, Ann. Phys.
{\bf 228} (1993), 1.

\bibitem{NTS} T. Nieuwenhuis, J. A. Tjon and Yu. A. Simonov,
contribution to the European Few-Body Conference XIV, Amsterdam,
23 -- 27 August 1993, p. 158.

\bibitem{LeDa} L. Di Leo and J. W. Darewych, Can. J. Phys.
{\bf 70} (1992) 412; {\bf 71} (1993) 365.

\bibitem{Saw} M. Sawicki, Phys. Rev. {\bf D 32} (1985), 2666.

\bibitem{Hil} J. R. Hiller, Phys. Rev. {\bf D 44} (1991), 2504.

\bibitem{WiHi} J. J. Wivoda and J. R. Hiller, Phys. Rev. {\bf D 47}
(1993), 4647.

\bibitem{Ji} C.-R. Ji, Phys. Lett. {\bf B 322} (1994) 389.

\bibitem{KLB} D.C. Khandekar, S. V. Lawande and K. V. Bhagwat:
Path-integral Methods and their Applications, World Scientific (1993).

\bibitem{Fock} V. Fock, Phys. Zeit. der Sowjetunion {\bf 12}
(1937), 404.

\bibitem{Nambu} Y. Nambu, Prog. Theor. Phys.{\bf 5} (1950), 82.

\bibitem{FeyQED} R. P. Feynman, Phys. Rev. {\bf 80} (1950), 440.

\bibitem{Schwi} J. Schwinger, Phys. Rev. {\bf 82} (1951), 664.

\bibitem{KaKt} A. I. Karanikas and C. N. Ktorides, Phys. Lett.
 {\bf B 275} (1992), 403; Int. J. Mod. Phys. {\bf A7} (1992), 5563.

\bibitem{McRe} D. G. C. McKeon and A. Rebhan, Phys. Rev. {\bf D 48}
(1993), 2891.

\bibitem{Str} M. J. Strassler, Nucl. Phys. {\bf B 385} (1992), 145.

\bibitem{ScSc} M. G. Schmidt  and Ch. Schubert, Phys. Lett.
{\bf B 331} (1994), 69.

\bibitem{TS} T. Nieuwenhuis and J. A. Tjon, preprint THU-95/16,
hep-ph/9506346, to appear in Phys. Lett. {\bf B}.

\bibitem{Schul} L. S. Schulman: Techniques and Applications of Path
Integration'', John Wiley (1981).

\bibitem{BDH} L. Brink, P. DiVecchia and P. Howe, Nucl. Phys.
{\bf B 118} (1977), 76

\bibitem{GrRy}I. S. Gradshteyn and I. M. Ryzhik: Table of Integrals,
Series and Products, Academic Press (1980).

\bibitem{FrGa} H. M. Fried and Y. M. Gabellini, Phys. Rev. {\bf D 51} (1995),
906 .

\bibitem{Light} M. J. Lighthill: Introduction to Fourier Analysis
and Generalised Functions, Cambridge University Press (1958).

\bibitem{Act} A. A. Actor, Fortschr. Phys. {\bf 41} (1993), 461.

\end{thebibliography}
\end{document}